 \documentclass[sigconf]{acmart}

\AtBeginDocument{%
  }

\setcopyright{none} 
\copyrightyear{2026}
\acmYear{2026}
\acmDOI{XXXXXXX.XXXXXXX}

\acmConference[MICRO 2026]{The 58th IEEE/ACM International Symposium on Microarchitecture}{October 31--November 04, 2026}{Athens, Greece}

\acmISBN{978-X-XXXX-XXXX-X/XX/XX}

\usepackage[]{hyperref}
\usepackage{xspace}

\usepackage{svg}
\usepackage{framed}
\usepackage{multirow}
\usepackage{url}
\usepackage{balance}
\usepackage{booktabs}
\usepackage{listings}
\usepackage{caption}
\usepackage{upquote}
\usepackage{xcolor}
\usepackage{comment}
\usepackage{subfigure}
\usepackage{mathtools}
\usepackage{ragged2e}
\usepackage{tikz}
\usepackage{graphicx}
\usepackage{listings}
\usepackage{xcolor}
\usepackage{mathtools}
\usepackage{booktabs}

\usepackage{dblfloatfix}    
\usepackage{svg}
\usepackage{framed}
\usepackage{multirow}
\usepackage{url}
\usepackage{balance}
\usepackage{booktabs}
\usepackage{listings}
\usepackage{caption}
\usepackage{upquote}
\usepackage[dvipsnames]{xcolor}
\usepackage{comment}
\usepackage{enumitem}
\usepackage{subfigure}
\usepackage{mathtools}
\usepackage{ragged2e}
\usepackage{tikz}
\usepackage{graphicx}
\usepackage{pifont}
\usepackage[table,xcdraw]{xcolor}

\usepackage{svg}
\usepackage{color}

\usepackage{siunitx}
\usetikzlibrary{arrows.meta, positioning, shapes.misc, calc, fit, backgrounds}
\usetikzlibrary{arrows.meta, positioning, shapes.misc, shapes.multipart, calc, fit, backgrounds}
\usetikzlibrary{arrows.meta, positioning, shapes.misc, shapes.geometric, shapes.multipart, fit, calc, backgrounds}

\definecolor{darkgreen}{rgb}{0,1,0}
\definecolor{codegreen}{rgb}{0,0.6,0}
\definecolor{codegray}{rgb}{0.5,0.5,0.5}
\definecolor{codepurple}{rgb}{0.58,0,0.82}
\definecolor{backcolour}{rgb}{0.95,0.95,0.92}
\newcommand{\cmark}{\textcolor{ForestGreen}{\ding{51}}}%
\newcommand{\xmark}{\textcolor{red}{\ding{55}}}%

\lstdefinestyle{mystyle}{
    backgroundcolor=\color{backcolour},
    commentstyle=\color{codegreen},
    keywordstyle=\color{magenta},
    numberstyle=\tiny\color{codegray},
    stringstyle=\color{codepurple},
    basicstyle=\ttfamily\footnotesize,
    breakatwhitespace=false,
    breaklines=true,
    captionpos=b,
    keepspaces=true,
    numbers=left,
    numbersep=5pt,
    showspaces=false,
    showstringspaces=false,
    showtabs=false,
    tabsize=2
}
\setcopyright{cc}
\acmConference{}{}{}
\acmBooktitle{}
\acmDOI{}
\acmISBN{}

\newcommand\encircle[1]{%
	\tikz[baseline=(X.base)] 
	\node (X) [draw, shape=circle, inner sep=0, fill=black, text=white, font=\small] {\strut #1};%
}

\newcommand{\papertitle}{Space-Control\xspace}

\newcommand{\context}{\texttt{HWPID}, \texttt{BASE\_P}\xspace}
\newcommand{\id}{ContextID\xspace}

\begin{document}


\title{\papertitle: Process-Level Isolation for Sharing CXL-based Disaggregated Memory}

\author{Kaustav Goswami}
\email{kggoswami@ucdavis.edu}
\affiliation{%
  \institution{University of California, Davis}
  \city{Davis}
  \state{California}
  \country{United States}
}

\author{Sean Peisert}
\email{sppeisert@lbl.gov}
\affiliation{%
  \institution{Lawrence Berkeley National Laboratory}
  \city{Berkeley}
  \state{California}
  \country{United States}
}

\author{Venkatesh Akella}
\email{akella@ucdavis.edu}
\affiliation{%
  \institution{University of California, Davis}
  \city{Davis}
  \state{California}
  \country{United States}
}

\author{Jason Lowe-Power}
\email{jlowepower@ucdavis.edu}
\affiliation{%
  \institution{University of California, Davis}
  \city{Davis}
  \state{California}
  \country{United States}
}

%
\renewcommand{\shortauthors}{K. Goswami, \textit{et al.}}

\begin{abstract}
Memory disaggregation via CXL enables multi-host resource sharing.
However, existing CXL sharing mechanisms enforce coarse-grained, host-level permissions only, leaving isolation to the operating system.
Today, virtual memory enables process-level isolation on a host and CXL enables host-level isolation.
This creates a critical security gap: the absence of process-level memory isolation in shared disaggregated memory.
We present \papertitle, an architectural abstraction that introduces a cross-host identity primitive to enforce confidentiality and integrity.
We decouple authorization from the untrusted OS using a hardware-rooted validation engine (SPACE) to establish immutable process identity and a Permission Checker at the memory egress point for fine-grained permission validation.
Our design supports 127 concurrent processes across 255 hosts with only 1.56\% storage overhead.
Cycle-level evaluation using gem5 + SST shows that Space-Control incurs a minimal 3.3\% performance penalty with a modest 16 KiB cache, providing a practical and scalable foundation for secure, process-level memory disaggregation.

\end{abstract}


\maketitle 

\pagestyle{plain} 

\section{Introduction}
    \label{sec:intro}

    The growing demand for memory in modern data-intensive applications has driven the industry toward memory disaggregation~\cite{og-dismem}.
    Disaggregation decouples compute and memory resources and connects them via high-speed fabrics.
    Standards such as the Compute Express Link (CXL)~\cite{cxl-main, cxl-3.1-spec, cxl-4-spec, cxl_sharing_2024} enable multiple systems, called \textit{hosts}, to access memory on a remote memory-only \textit{device}.
    This improves utilization and reduces redundancy~\cite{pond}.
    The CXL 3.0~\cite{cxl-3.1-spec} standard supports memory sharing at the cache-line granularity across hosts for the first time, making it a promising foundation for large-scale deployments.
    
    Current CXL-based sharing mechanisms provide coarse-grained, host-level access control: \textit{once a host is authorized, all processes on that host can access the shared region.}
    This violates the \textit{principle of least privilege}~\cite{lampson}.
    The \textit{all-or-nothing} model leaves intra-host isolation to the virtual memory.
    Virtual memory provides process-level isolation locally, which is invisible across hosts.
    Since memory is shared across multiple hosts, there is no cross-host user or process visibility.
    When coupled with a compromised kernel~\cite{google-zero,spectre,rowhammer,intel-patent}, this gap can escalate privileges and access sensitive data, creating a significant security risk~\cite{deact,cosm,cheri-flexible}.

    Memory isolation has been largely addressed by trusted execution environments (TEEs) on traditional single hosts~\cite{intel-sgx,intel-tdx,amdsev,cxl-4-spec,cxl-3.1-spec,data-enclaves,ayaz-thesis,elasticlave}.
    TEEs provide enclave, a critical isolation primitive, that functions without trusting the operating system (OS).
    However, CXL introduces a new challenge: sharing memory without cross-host visibility.

    Commercial TEE designs like Intel SGX~\cite{intel-sgx}, Intel TDX~\cite{intel-tdx}, AMD SEV~\cite{amdsev}, CXL-TEE (TSP)~\cite{cxl-4-spec,cxl-3.1-spec} \textit{etc.}, cannot share memory between enclaves~\cite{elasticlave,data-enclaves,cxl-4-spec,cxl-3.1-spec}.
    There have been research-based designs for sharing data between enclaves~\cite{elasticlave,data-enclaves}.
    However, these designs are for single-host systems only, as the security monitor, the hardware enforcer, operates locally, with no cross-host coordination. 
    Furthermore, TEEs introduce notable performance overheads, reduce deployment flexibility, and expand the trusted computing base (TCB)~\cite{ayaz-ipdps}.
    
    Exploring different memory isolation designs led us to ask a fundamental question, \textit{``What is the minimum addition needed for shared disaggregated memory isolation?''}
    Hence, we revisit OS-based memory isolation to understand and address the missing gap.
    On a single-host system, the OS marks page table entries for sharing, and the TLB enforces permissions in the hardware. 
    The POSIX shared memory API~\cite{shmem-linux} simplifies inter-process shared memory and management using usergroups.
    However, the missing gap is the lack of cross-host visibility.
    This, therefore, leads us to the central research question in this paper:
    \begin{quote}
        ``How can we ensure isolation in shared disaggregated memory
        that enforces the principle of least privilege and operates
        independently of the OS with minimal overhead?''
    \end{quote}

    Prior works have emphasized \emph{horizontal} access control, \textit{i.e.} host-level permissions~\cite{deact,cxl-3.1-spec,cxl-4-spec}, whereas shared disaggregated memory (SDM) also needs \emph{vertical} access control, \textit{i.e.} per-process or per-virtual machine isolation across hosts.
    Vertical control requires a cross-host notion of \textit{who} is issuing a memory access, independent of the host OS.
    Therefore, we introduce \papertitle, a new security abstraction, that adds architectural cross-host \textit{identity primitive} for isolation.
    The vertical abstraction in our work is realized through two key architectural components: \textit{(a)} a new host-side validation hardware, and, \textit{(b)} a permission checker with a storage-efficient table.
    The former establishes a hardware-rooted identity at every context switch.
    The latter validates each load/store (LD/ST) at the memory egress point for arbitrary memory ranges, enabling fine-grained access control.
    The goal is to have a flexible design, meaning that the identity primitive shall be user-defined (\textit{e.g.} \textit{Linux processes}, \textit{capabilities}~\cite{cheri-paper}, \textit{virtual machines}, or, \textit{enclaves}) for cross-host sharing.

    In this work, we show the compatiblity of \papertitle with the existing virtual memory infrastructure by enforcing isolation defined by a Linux process. 
    A process is the smallest OS abstraction that issues LD/ST under its own distinct address space.
    \papertitle registers and authenticates address spaces defined by a user process at the host using a secure process attribute context engine (SPACE), instead of the OS.
    Further, we leverage the CXL fabric manager (FM) for managing cryptographic keys and permissions, making it the root-of-trust.

    
    Disaggregated memory workloads~\cite{famfs,kvstore,cxl-graphs} is an active research topic today.
    These workloads usually partition their data across local and remote memory.
    The CXL specification assigns coarse-grained regions (256 MiB) to hosts for sharing memory~\cite{cxl-3.1-spec,cxl-4-spec}.
    A lack of a fine-grained access control has been the hindrance to the wider adoption of shared CXL memory.
    Tigon~\cite{kvstore}, a CXL-based key-value store software, explicitly calls out the lack of a fine-grained access controls in CXL.

    The core difficulty is balancing fine-grained authorization with scalability and efficiency.
    Tagging every 4 KiB page of a 1 TiB memory region with permissions (2 bits) per host (up to 256) and per process (up to 128) requires 2 TiB of metadata (200\% overhead).
    Therefore, na\"ive lookup table designs become infeasible, and permission checks risk becoming performance bottlenecks.
    Prior research like Mondrian~\cite{mmp} trades off storage overhead for higher lookup latency on a single host using sorted tables.
    However, such tables must be replicated for cross-host visibility. 

    To address the storage overhead problem, \papertitle stores cache-line-sized permissions as a range-based table in SDM where each entry carries compact host/process-context bitmasks, yielding a small fixed metadata footprint.
    This reduces the total storage overhead to a fixed 1.56\% of the total memory capacity when 255 hosts share memory of 4 KiB pages across 127 processes on each host.
    
        
    Our design delivers hardware-backed isolation with space-efficient metadata, addressing both the flexibility and granularity requirements of SDM.
    \papertitle's evaluation on GAPBS~\cite{gapbs} shows a marginal 3.3\% performance overhead with a modest 16 KiB permission cache.
The contributions of this paper are:

\begin{itemize}
    \item We identify a critical isolation gap in Shared Disaggregated Memory (SDM).
        While current systems provide process-level isolation (via the OS) and host-level isolation (via CXL), they lack process-level isolation for SDM.
    \item We propose \papertitle, a new security abstraction.
        We preserve process identity primitive across hosts, that enables SDM isolation even if the OS is compromised.
        This architecture includes a hardware validation module and an access-control permission checker.
    \item We present the hardware design and provide a quantitative evaluation of its performance and space overheads.
    \item Our design reduces access-control metadata overhead through hardware-software co-designed permission management.
\end{itemize}

    The rest of the paper is organized as follows.
    \S\ref{sec:background} explains the background.
    \S\ref{sec:method} provides an overview of the problem and the solution.
    \S\ref{sec:impl} further dives into the implementation details.
    We disucss the implications of \papertitle in \S\ref{sec:discussion}.
    The following sections discusses the evaluation methodology and results. 
    \S\ref{sec:related} explains related works.
    The work is concluded in \S\ref{sec:conclusion}.

\section{Background} 
    \label{sec:background}
    To contextualize the contributions of \papertitle, we first outline the limitations of existing single-host and disaggregated access control mechanisms.


    On a single-host system, virtual memory manages per-process page tables.
    The hardware (TLB/MMU) enforces translation and per-page \textit{W/R/X} permissions~\cite{osbook}.
    A process cannot access another's pages unless they are explicitly shared.
    Address space identifier (ASID; ARM) or process context identifier (PCID; X86) tags TLB entries to distinguish address spaces and reduce TLB shootdowns.
    They aid performance but do not provide authorization.

    Intel's PASID/SVA~\cite{linux-pasid,intel-pasid} binds I/O accelerators to a process' virtual address space for device-initiated DMA via the IOMMU.
    This is orthogonal to CPU-originated LD/ST. 
    OS-managed mechanisms such as Border-Control~\cite{bordercontrol}, MPK~\cite{intel-mpk} and ARM MTE~\cite{arm-mte} provide intra-host memory protection but assume a trusted kernel. 
    Mondrian memory protection introduces \textit{domains} as a set of processes with identical permissions, defined by the OS.
    These domains are sorted and stored as entries in a permission table.


    Disaggregated memory~\cite{og-dismem} decouples compute from memory and lets hosts map SDM ranges over a fabric~\cite{mem-disaggregation}.
    The fabric manager (FM) manages the set of hosts sharing memory on a set of devices.
    Prior efforts include OpenCAPI~\cite{opencapi} and Gen-Z~\cite{genz}, but current deployments center on CXL~\cite{cxl-main, cxl-3.1-spec, cxl-4-spec, cxl_sharing_2024,pond}.
    The critical difference for our work is the \emph{translation and authorization path}.
    
    Gen-Z introduced a host-side zMMU and two-stage translation (virtual address (VA) $\!\to\!$ remote virtual address $\!\to\!$ physical address (PA)), which enabled designs like DeACT~\cite{deact} to decouple translation from access-control.
    In contrast, CXL removes the host-side zMMU and reuses the CPU's TLB to access SDM memory directly.
    With CXL~3.0, multiple hosts may share the same physical memory range~\cite{cxl-3.1-spec} (called \textit{sharing}).
    To avoid the OS zeroing-out the bits, an SDM range is mapped as a direct access (DAX) region, similar to the heterogeneous programming model~\cite{famfs,hdcs-paper}.
    However, authorization is \emph{host-level} and coarse (dynamic capacity region granularity) under the FM. 


    
    Prior works~\cite{deact,cheri-rackscale} either target the Gen-Z model or rely on a trusted OS for isolation~\cite{legoos}.
    CXL's TEE design, called Transport Security Protocol (TSP) extends existing TEEs at the host, and provides enclave support for memory pooling, but not for sharing~\cite{cxl-3.1-spec, cxl-4-spec}. 
    CHERI-style capability~\cite{cheri-sp,cheri-paper,cheri-rackscale} systems offer principled, fine-grained authority tracking~\cite{cheri-flexible}, but adoption requires ISA/compiler/OS changes across all hosts.
    CHERI uses 128 bit pointers to maintain capability per host, maintaining the metadata across 256 hosts takes 12.5\% more storage ($256 \times 1b$ per 16 Bytes).
    In addition, programs need to be rewritten to understand \textit{capabilities-aware} LD/ST instructions~\cite{cheri-paper}.

\section{\papertitle}
    \label{sec:method}
In this section, we present the design of \papertitle.
We first define our adversarial threat model and design requirements before detailing the core architectural abstraction and the overview of the components that realize this abstraction.

    \subsection{Threat Model}
        \label{sec:threat_model}
        \paragraph{Principle and Scope.}
            \papertitle's trust model follows the principle of least privilege~\cite{lampson,polp}. 
            Only address spaces defined by a process, that require access to the remote (SDM) memory, should be allowed to access the same.
            We provide \textit{confidentiality} to local and remote pages of such process' data.
            Further, we provide \textit{integrity} to all remote pages and detect any intrusion by the OS.

        \paragraph{System setting.}
            We consider a HPC-like deployment where several multi-tenant hosts share SDM via CXL.
            Users directly interact with the host OS via user processes without virtual machines. 
            Applications interact with SDM through user processes that map regions (\textit{e.g.}, DAX).
            Each process has both local pages and remote pages.

        \paragraph{Trusted.}
            We trust the user process (its code and keys) that is granted SDM access, and the additional hardware we propose.
            We assume on-chip and CXL hardware to function correctly.

        \paragraph{Untrusted.}
            The OS (including the kernel) is untrusted in the kernel mode. 
            Our only kernel change is a narrow extension that enables the OS to understand the security abstraction and SDM ranges; it contains no authorization logic. 
            If this change is absent, SDM remains unmapped and the hardware permission checker rejects all SDM requests.
            Other user or kernel processes on the host may collude with the OS to break confidentiality.
            Off-chip components such as local DRAM modules and other external devices are untrusted.
        
        \paragraph{Attacker Capabilities.}
            We consider a powerful adversary capable of controlling the OS.
            OS-level vulnerabilities can escalate privileges and override permissions~\cite{deact,spectre,meltdown,google-zero}.
            The attacker can manipulate page tables, virtual memory mappings, replay or forge identity, and invoke privileged instructions.
            Physical attacks and side-channels on the hardware are out of scope.

    \begin{table}[t]
        \centering
        \resizebox{0.48\textwidth}{!}{%
            \begin{tabular}{c||cccc||}
                \hline
                \rowcolor[HTML]{C0C0C0} 
                \cellcolor[HTML]{C0C0C0} & \multicolumn{4}{c|}{\cellcolor[HTML]{C0C0C0}\textbf{Requirements}} \\ \cline{2-5} 
                \rowcolor[HTML]{C0C0C0} 
                \multirow{-2}{*}{\cellcolor[HTML]{C0C0C0}\textbf{Work}} & \multicolumn{1}{c|}{\cellcolor[HTML]{C0C0C0}\textbf{\begin{tabular}[c]{@{}c@{}}R1: Principle of\\ Least Privilege\end{tabular}}} & \multicolumn{1}{c|}{\cellcolor[HTML]{C0C0C0}\textbf{\begin{tabular}[c]{@{}c@{}}R2: Small TCB\end{tabular}}} & \multicolumn{1}{c|}{\cellcolor[HTML]{C0C0C0}\textbf{\begin{tabular}[c]{@{}c@{}}R3: Low Access \\ Control Metadata\end{tabular}}} & \textbf{\begin{tabular}[c]{@{}c@{}}R4: Memory\\ Sharing\end{tabular}} \\ \hline \hline
                Mondrian~\cite{mmp} & \multicolumn{1}{c|}{\cmark} & \multicolumn{1}{c|}{\xmark} & \multicolumn{1}{c|}{\xmark} & \cmark \\ \hline
                Border Control~\cite{bordercontrol} & \multicolumn{1}{c|}{\cmark} & \multicolumn{1}{c|}{\xmark} & \multicolumn{1}{c|}{\xmark} & \cmark \\ \hline
                Intel MPK~\cite{intel-mpk} & \multicolumn{1}{c|}{\cmark} & \multicolumn{1}{c|}{\xmark} & \multicolumn{1}{c|}{\cmark} & \cmark \\ \hline
                ARM MTE~\cite{arm-mte} & \multicolumn{1}{c|}{\cmark} & \multicolumn{1}{c|}{\xmark} & \multicolumn{1}{c|}{\cmark} & \cmark \\ \hline
                CHERI-HW~\cite{cheri-paper} & \multicolumn{1}{c|}{\cmark} & \multicolumn{1}{c|}{\cmark} & \multicolumn{1}{c|}{\xmark} & \cmark \\ \hline
                DeACT~\cite{deact} & \multicolumn{1}{c|}{\xmark} & \multicolumn{1}{c|}{\xmark} & \multicolumn{1}{c|}{\cmark} & \cmark \\ \hline
                LegoOS~\cite{legoos} & \multicolumn{1}{c|}{\cmark} & \multicolumn{1}{c|}{\xmark} & \multicolumn{1}{c|}{\xmark} & \cmark \\ \hline
                Intel SGX~\cite{intel-sgx} & \multicolumn{1}{c|}{\cmark} & \multicolumn{1}{c|}{\cmark} & \multicolumn{1}{c|}{\xmark} & \xmark \\ \hline
                AMD SEV~\cite{amdsev} & \multicolumn{1}{c|}{\cmark} & \multicolumn{1}{c|}{\xmark} & \multicolumn{1}{c|}{\cmark} & \xmark \\ \hline
                CXL 3.0+~\cite{cxl-3.1-spec,cxl-4-spec} & \multicolumn{1}{c|}{\xmark} & \multicolumn{1}{c|}{\xmark} & \multicolumn{1}{c|}{\cmark} & \cmark \\ \hline
                CXL TEE (TSP)~\cite{cxl-3.1-spec} & \multicolumn{1}{c|}{\cmark} & \multicolumn{1}{c|}{\cmark} & \multicolumn{1}{c|}{\cmark} & \xmark \\ \hline
                \papertitle & \multicolumn{1}{c|}{\cmark} & \multicolumn{1}{c|}{\cmark} & \multicolumn{1}{c|}{\cmark} & \cmark \\ \hline \hline 
            \end{tabular}%
        }
        \caption{The table highlights that \papertitle is the only approach to the best of our knowledge, that achieves process-level isolation, OS independence, scalability, and efficiency simultaneously, while minimizing the TCB.}
        \label{tab:trusted_comparison}
    \end{table}
    \subsection{Requirements}
        \label{sec:requirements}
        Given the scope and the threat model, any given solution needs to have the following requirements:
        \begin{enumerate}[label=R\arabic*]
            \item It should enforce both horizontal and vertical access-control to ensure the \textit{principle of least privilege}.
            \item The solution needs to operate independently of the OS, so that isolation remains intact even if the OS is compromised, \textit{thereby minimizing the TCB}.
            \item The design must scale with \textit{low metadata footprint} and small per-access overhead. 
            \item Allow dynamic creation/teardown of \textit{shared} memory ranges and policy changes across processes. 
        \end{enumerate}

    Table~\ref{tab:trusted_comparison} summarizes how prior mechanisms from \S\ref{sec:background} meet these requirements and highlights how we address the remaining gap. 

    \subsection{From Requirements to Abstraction}
        \label{sec:gap}
        To satisfy the listed requirements, \papertitle decouples translation from authorization.
        This decoupling directly addresses the design constraints as follows:
        \paragraph{Identity Primitive and OS Independence (R2).} Since we extend the virtual memory infrastructure in this work, we therefore anchor the process identity in the hardware.
        We call it \textit{\id}, which is defined as (\context), where \texttt{HWPID} is a set of reserved PCID/ASID, and, \texttt{BASE\_P} is the pointer to the base of the page table (\texttt{CR3/SATP/TTBR}).
        The abstraction remains intact even if the host kernel is compromised.
        \paragraph{Enabling fine-grained permissions (R1).} The identity primitive allows the system to enforce permissions at the process level for arbitrary memory ranges, upholding the principle of least privilege.
        \paragraph{Optimizing Scalability and Performance (R3, R4).} By representing the identity primitive as space-efficient bit-vectors in a global permission table for seamless sharing, we minimize metadata storage and lookup latency.

    \subsection{System Overview and Key Insights}
        \label{subsec:overview}
        Like any security abstraction, we have three simple steps for enforcing SDM isolation: 
        \textit{(i)} \textit{authenticate} the current process,
        \textit{(ii)} \textit{tag} its memory requests with a hardware-backed identifier, and,
        \textit{(iii)} \textit{validate} each SDM access against a permission table.
        Towards this, \papertitle introduces three key components:
        \begin{enumerate}
            \item \textit{Secure Process Attribute Context Engine (SPACE), or the authenticator:}
                    A lightweight hardware module that authenticates processes using \id.
                    The engine monitors context switches and ensures that only validated \id can access the SDM.
                    The insight here is that SPACE shifts the permission enforcement outside the OS, thus eliminating the OS as a threat vector.
                    It offers minimal MMIO interface to the programmer.
                    In addition, it has logic for generating cryptographic labels, and, a counter.
            \item \textit{Permission Table, or the metadata:}
                    An array of permission entries that maps PA ranges; shared or exclusive; of the SDM to the set of authorized \id across hosts.
            \item \textit{Permission Checker, or the enforcer:}
                    An on-chip hardware unit placed after the last-level cache, before the local DRAM controller and the CXL downstream port, responsible for enforcing permissions on each LD/ST. 
                    It is integrated into the memory access pipeline that actively validates each SDM request against the \textit{permission table}.
        \end{enumerate}

        \papertitle makes a key observation in existing works: access control metadata exponentially increases  with increasing hosts and processes. 
        Therefore, we opt for a hardware-software co-design, explained in \S\ref{sec:table}.
        We minimally modify the kernel to reserve contiguous processor-context IDs (ASID/PCIDs (referred to a \texttt{HWPID} henceforth) instead of Linux PID) for SDM processes.
        \texttt{HWPIDs} are assigned by the SPACE instead of the OS.
        These are tagged with the PA for identity propagation across the fabric as the architectural manifestation of the identity primitive that we propose.
        The OS still remains untrusted as its correctness never depends on kernel behavior.
        If the OS is unmodified or these extensions are absent, SDM cannot be mapped or accessed, and the permission checker will always reject any SDM traffic.

        The FM serves as the global root-of-trust and manages cryptographic keys and global permissions.
        Its secret key ($K_{FM}$) is used to generate public authentication token labels ($L_{exp}$) to  validate a process identity at the host.

        \begin{figure}[t]
            \centering
        \includegraphics[width=0.8\columnwidth]{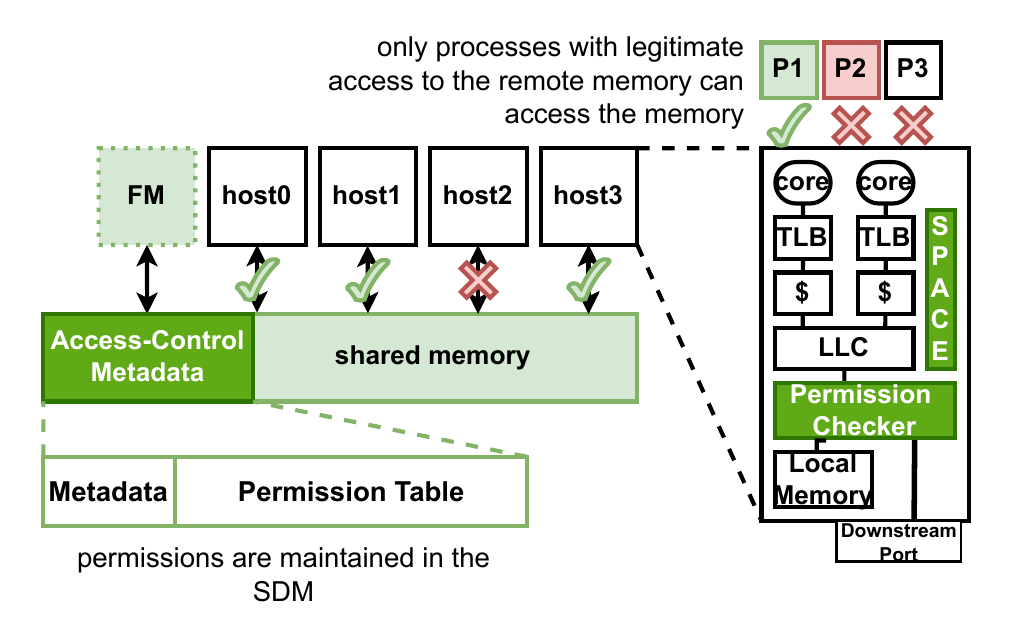}
            \caption{\papertitle architecture providing process-level isolation for shared disaggregated memory.
            With the addition of new trusted hardware, only the trusted process P1 on host3 is allowed access to the shared memory.
            In CXL, any process on host3 can access that data.
            }
            \label{fig:overview}
        \end{figure}

            \begin{figure}[!b]
                \centering
                \includegraphics[width=0.99\columnwidth]{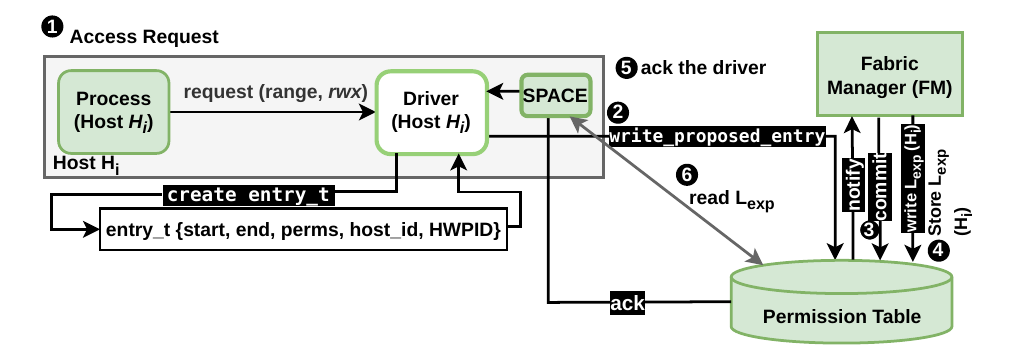}
                \caption{Process creation and permission setup workflow. 
                        A new process requests access to SDM, which proposes an entry to the metadata region.
                        The FM is automatically notified and decides approval, commits the proposed entry to the table and issues $L_{exp}$.
                }
                \label{fig:process_creation}
            \end{figure}
        Figure~\ref{fig:overview} shows the system-level architecture of \papertitle.
        While green in general represents trusted components, the darker shade represents additional hardware.
        The FM manages the permission table stored in SDM.
        The key difference between \papertitle and CXL is while CXL allows all processes and the kernel on host3 to access the shared memory, we only allow P1 to access that data. 
        Each host enforces access control through SPACE, which authenticates processes using the \id.
        The permission checker validates every LD/ST against the permission table before committing.
        Only trusted processes with legitimate permissions can issue LD/ST, while unauthorized processes are blocked. 



\section{Implementation Details}
    \label{sec:impl}

    \subsection{The Lifecycle of a Secure Identity}
        \label{subsec:flow}
        Now that we have given an overview of the design and selected hardware process context (\id $=$ (\context)) as the identity primitive, we explain the work flow in three distinct phases:
        \textit{(a)} process creation, binding and registration,
        \textit{(b)} runtime protection, and,
        \textit{(c)} dynamic updates.

        \subsubsection{Identity Binding and Registration}
            \label{sec:creation}
    
            Figure~\ref{fig:process_creation} illustrates the process creation and permission request workflow in \papertitle.
            The goal is to bind a software process to a hardware-enforced identifier (\context) without OS intervention.
            Towards this, the user registers their process as \textit{trusted} via SPACE using MMIO doorbells instead of the OS, and, sets up permissions accordingly. 
            
            Action~\encircle{1} requests a range of memory to be mapped into a trusted process' address space, running on the $i^{th}$ host.
            The user is responsible for requesting the right permissions.
            This can be achieved either using a \textit{middleware} or an \textit{upper layer protocol} as mentioned by prior works~\cite{hdcs-paper}.
            We use Linux's userspace I/O (UIO)~\cite{uio-kernel} driver, running completely in the user-space, for the same.

            SPACE maintains a list of free HWPIDs assigned via an MMIO doorbell (\texttt{GET\_NEXT\_PID()}).
            The driver assigns a HWPID from the free list. 
            This breaks the dependence on the OS for a process identifier.
            The permission table that records all permission entries is stored in the SDM.
            A new entry (\texttt{entry\_t}) is written into the dedicated section of the table (Action~\encircle{2}). 

            The FM is automatically notified about the update and decides whether to approve the request.
            If the entry is approved, the FM commits the entry (Action~\encircle{3}) and optimizes the table, if needed.
            The FM immediately generates a cryptographic public label ($L_{exp}$) using its private key (K\textsubscript{FM}) for authentication and stores it into the permission table (Action~\encircle{5}).
            In addition, it responds to the host with the public label (Action~\encircle{6}), which is intercepted by SPACE for cryptographic label comparison.
            The user process is now assumed to be trusted and the execution continues.

            
        \subsubsection{Runtime Protection and Identity Propagation}
            \label{sec:runtime}
    
            \begin{figure}[!b]
                \centering
                \includegraphics[width=0.99\columnwidth]{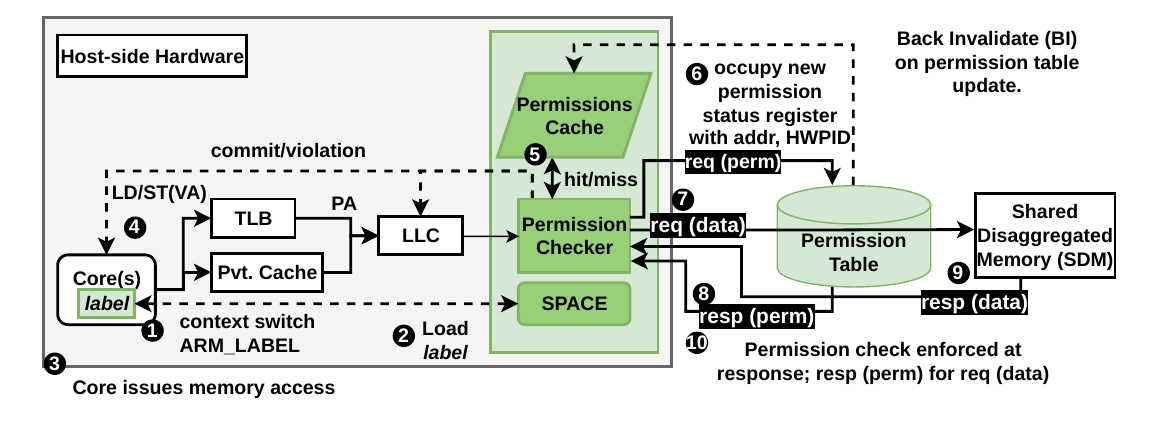}
                \caption{Flow of remote memory access (LD/ST) of a virtual address (VA) in \papertitle.
                SPACE authenticates the \id, and the permission checker enforces access-controls.
                }
                \label{fig:process_runtime}
            \end{figure}
            We now have a trusted and registered process.
            However, multi-tenant OS has other processes running in the system.
            Therefore, we need to validate our trusted process at every context switch.
            Figure~\ref{fig:process_runtime} depicts the runtime protections provided by \papertitle.

            When a new process is context switched (Action~\encircle{1}), the SPACE logic automatically reads the \texttt{BASE\_P} (\texttt{CR3/SATP/TTBR}) and the \texttt{HWPID} (\texttt{pcid/ASID}) registers. 
            Based on the current context, the trusted user process calls MMIO doorbell \texttt{ARM\_LABEL} to generate a host-side label ($L_{host} \leftarrow \id$) from the user-space.
            SPACE generates $L_{host}$ \textit{iff} \texttt{ARM\_LABEL} is invoked from user-space eliminating the possibility of the kernel falsely trying to validate a different \id.

            
            $L_{host}$ is loaded into a shadow register called the \textit{label register} (Action~\encircle{2}).
            Authentication succeeds if the current label $L_{host}$ matches with the public label from the FM \textit{i.e.} $L_{exp}$. 
            To avoid OS traps, kernel isolation and stale labels, the shadow register is automatically unset if the core's protection ring is anything other than the user-space.
            
            Trusted-process-generated PA (local or remote) (Action~\encircle{4}) are tagged with the HWPID bits (identity-tagged LD/ST (Action~\encircle{5})).
            We call these HWPID bits the authentication bits (\textit{A-bits}), as these are the architectural manifestation of the identity primitive, and ensure R1 is satisfied at all times.
            
            The permission checker checks the tagged A-bits for all LD/ST by sending a permission request (Action~\encircle{6}) to the table.
            The corresponding permission response has a range and the HWPIDs of authenticated processes.

            While LDs are sent together with the permission request (Action~\encircle{7}), STs are stalled until the response arrives.
            For higher performance, permission on an SDM load is enforced at response time (Action~\encircle{10}).
            On the other hand, for security, permission on an SDM store is enforced at the request time (Action~\encircle{7}). 
            An interrupt is generated, and handled by OS, if there is an access violation.

            To ensure the confidentiality of local memory pages of the trusted process, a memory encryption engine is used.
            The OS cannot forge identity-tagged LD/ST.
            Even if a malicious OS tampers with the trusted process's page-table entries, any data it can exfiltrate from local memory is ciphertext, thus maintaining \textit{confidentiality}.

        \subsubsection{Revocation and Update Propagation}
            We rely on CXL's back invalidates (issued via \texttt{BISnp}) to propagate updates to the permission table.
            When an update is committed to the table by the FM, every host automatically receives a \texttt{BISnp}, ensuring that the permission is invalid even if the entry is cached in the permission cache.
            The FM updates the $L_{exp}$ for the same host, ensuring global transparency. 
            The UIO driver is responsible for calling SPACE doorbells for cleaning up HWPIDs.
            Permission entries with no hosts are automatically cleaned up by the FM.
            

    \subsection{\papertitle Components}
        We now explain each component within the workflow mentioned in \S\ref{subsec:flow}, describing their individual microarchitectures.
        \subsubsection{Secure Process Attribute Context Engine (SPACE)}
\begin{figure}[t]
                \centering
                \includegraphics[width=0.9\columnwidth]{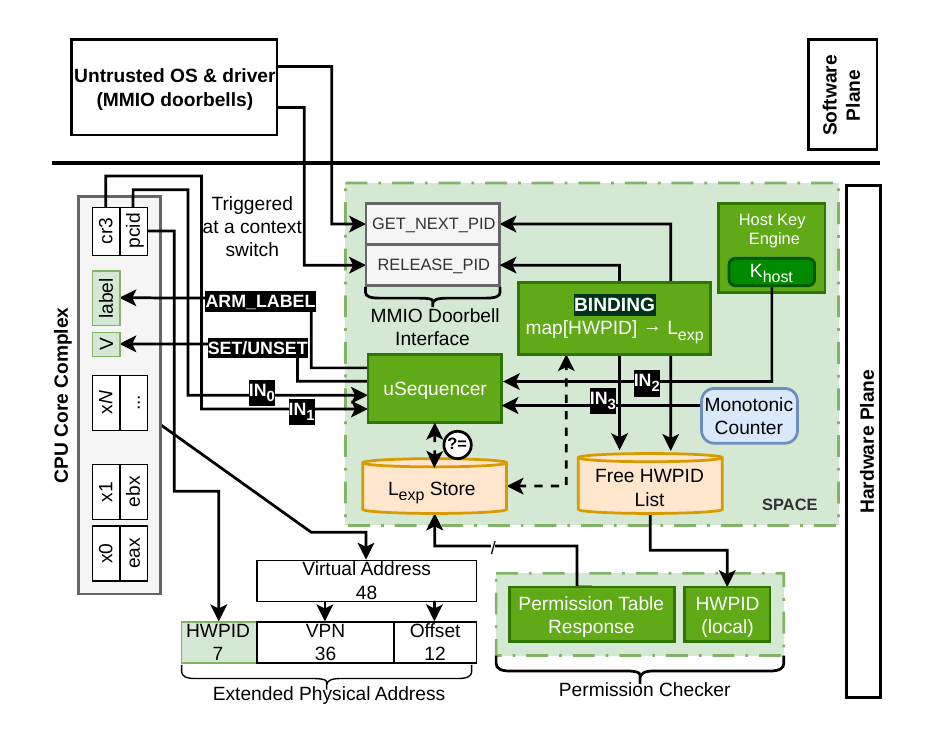}
                \caption{The microarchitecture of the SPACE.
                        It stores the host's secret key ($K_{host}$), FM's public label ($L_{exp}$) and a free \texttt{HWPID} list.
                        There is additional logic to generate $L_{host}$.
                        Further, SPACE provides minimal MMIO doorbells available to the OS for HWPID management.
                }
                \label{fig:space_micro}
            \end{figure}

            SPACE, shown in Figure~\ref{fig:space_micro}, is the hardware root-of-trust for process authentication.
            It binds a process with its hardware-rooted identity (Linux PID $\rightarrow$ \id).
            It stores the $L_{exp}$ and a list of free \texttt{HWPIDs}.
            SPACE ensures that only processes explicitly authorized by the FM can access SDM.
            It has a \texttt{$\mu$Sequencer}, that monitors context switches and generates cryptographic labels.
            Further, it provides MMIO doorbells for the OS system to interact.

            To establish \id, \papertitle uses two complementary labels:
            \textit{(a)} the FM-issued public label ($L_{exp}$), and
            \textit{(b)} the SPACE generated label ($L_{host}$) for the currently running process.
            These labels are generated using standard MAC primitive (\textit{e.g.} HMAC-SHA-256~\cite{hmac,specific} or AES-CMAC~\cite{aes}).
            There are hardware implementations of such algorithms~\cite{hwimpl}.

            During the process creation phase (\S\ref{sec:creation}), the FM generates $L_{exp}$ that acts as an authorization token that binds a process (on a specific host) to the memory range(s) it may access.
            Conceptually,
            \begin{equation}
                L_{exp} = MAC_{K_{FM}} (host\_id, HWPID, BASE\_P, range)
            \end{equation}
            Where $K_{FM}$ is the FM's secret key.
            SPACE stores the authorization label for subsequent checks.

            On the other hand, the SPACE-generated host-side label $L_{host}$ is generated and stored in the \textit{label register} \textit{iff} the authorized user-space process invokes \texttt{ARM\_LABEL} at a context switch (\S\ref{sec:runtime}).
            $L_{host}$ represents the \id of the process currently running on that core.
            Freshness of $L_{host}$ is ensured via a monotonic counter. 
            $L_{host}$ is given by:
            \begin{equation}
                    L_{host} = MAC_{K_{host}} (BASE\_P, HWPID, ctr)
            \end{equation}
            Where $K_{host}$ is the host's secret key known to the SPACE, and ctr is the monotonic counter that advances on each context switch per core.

            In summary, on a context switch, SPACE:
            \begin{enumerate}
                \item Derives the current process' \id.
                \item Generates $L_{host}$ and stores it in the \textit{label register} for that specific core.
                \item Compares the two labels to match the predicate. 
            \end{enumerate}
            Authenticated processes will always generate identity-tagged LD/ST.
            This enables per-access permission enforcement downstream and prevents synonyms in the memory. 

            All identity construction and label checks occur in trusted hardware.
            The OS cannot mint identities or forge labels.
            The monotonic counter ensures that $L_{host}$ is tied to exactly one context switch, which prevents replay attacks~\cite{tocttou}.
            This ensures OS independence (R2) and upholds the principle of least privilege (R1) as authorization tokens bind processes to specific SDM ranges.


            \begin{figure}[t]
                \centering
                \includegraphics[width=0.99\columnwidth]{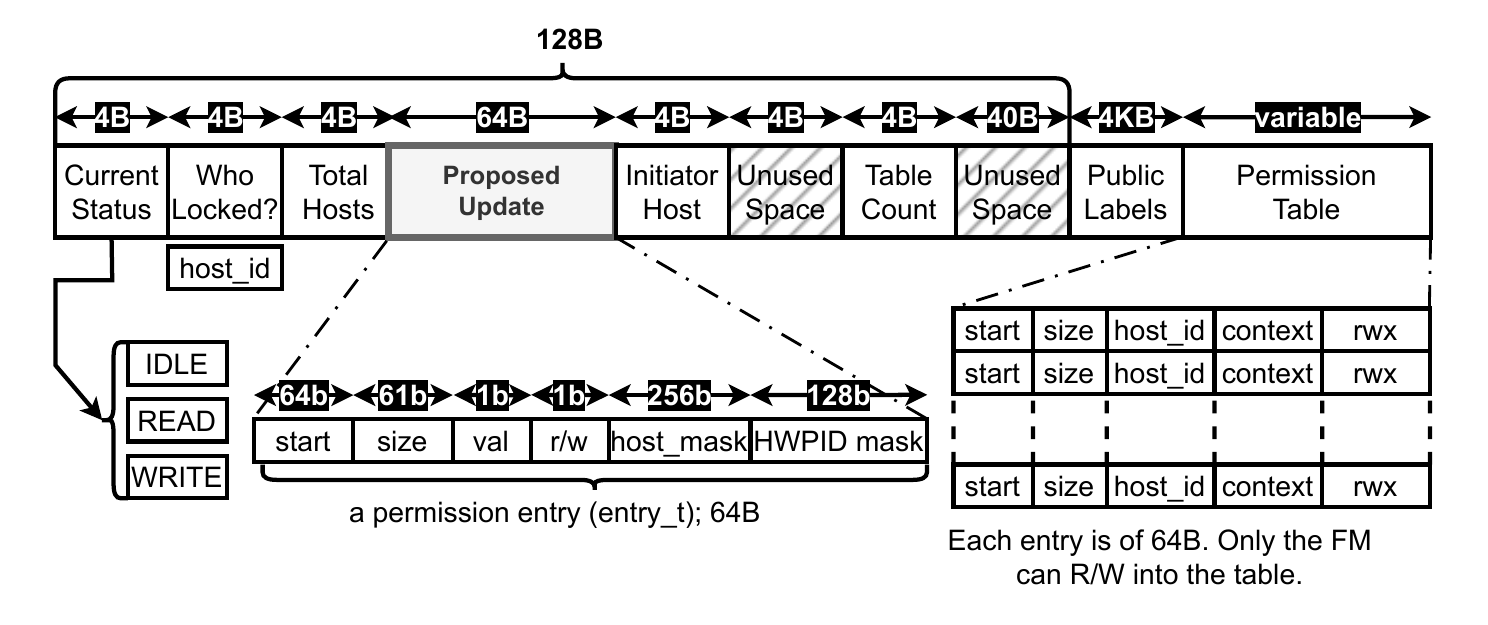}
                \caption{Layout of the permission table stored in the SDM.
                        The permission table starts at an offset 128 B (cache-line aligned).
                        It stores sorted permission entries (entry\_t) keyed by start address.
                }
                \label{fig:permission_table}
            \end{figure}

        \subsubsection{Permission Table}
            \label{sec:table}
            The permission table (Figure~\ref{fig:permission_table}) stores access control metadata per process per host.
            Each entry on the table has a start, a size, and, permission attributes.
            It functions as an optimized reverse mapping table, inspired by directory-based caching protocols, that maps a PA to a host and a process within the host.
            The metadata section of the table allows hosts to add new updates to the table.
            The rest of the table with all the permission entries is only accessible to the FM.

            The organization of the permission table is central to ensuring scalability and satisfying requirement R3.
            The primary challenge is maintaining unique pairs of (host, process identifier) across all participating hosts.
            To achieve this, we maintain two representations of the HWPIDs, implemented as bit vectors, authorized to access the SDM. 
            The local copy, denoted by $HWPID_{local}$, contains the HWPIDs assigned to trusted processes on the local host.
            The permission table copy, $HWPID_{global}$, represents the union of all trusted processes across all hosts and is defined as $HWPID_{global} = \bigcup _{i} HWPID_{local_i}$.
            The intersection $HWPID_{local} \bigcap HWPID_{global}$ validates the right process at the right host.

            The permission table is organized as a sorted table based on the start addresses, similar to prior works like Mondrian~\cite{mmp}.
            This allows flexibility in defining protections for arbitrary memory ranges instead of fixed page lengths. 

        \subsubsection{Permission Checker}

\begin{figure}[t]
                \centering
                \includegraphics[width=0.9\columnwidth]{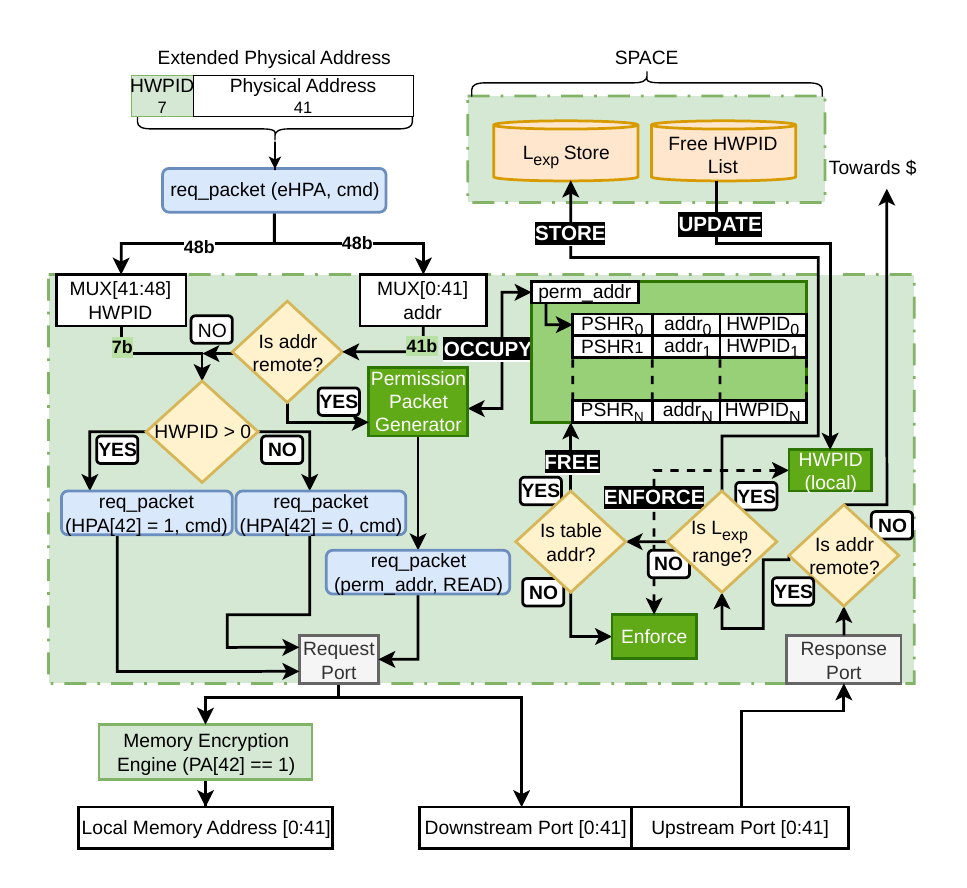}
                \caption{The microarchitecture of the permission checker.
                        The diagram shows the flow of a 41 bit PA (48 bit extended PA).
                        While every identity-tagged SDM LD/ST is validated against the permission table, local pages are encrypted to ensure confidentiality.
                }
                \label{fig:permission_micro}
            \end{figure}

            The permission checker, shown in Figure~\ref{fig:permission_micro} is designed in tandem with the previous components.
            It is placed on-chip of a host, after the last-level cache. 
            The checker validates whether the trusted process' identity-tagged addresses have access rights to a given PA at all times. 
            
            At runtime, every SDM LD/ST is associated with a corresponding permission entry in the table that encodes the range and permission attributes.
            The checker fetches the same entry from the permission table.
            Homonyms never appear in our design as PA are used for range checks.


            The lookup latency is proportional to the number of permission entries.
            To amortize the lookup latency and avoid round-trips on every remote memory reference, we add a small fully-associative permission cache.

            To track each outgoing request and its corresponding permission, we use a limited number of permission status holding registers.
            Permissions are enforced at the request-end for STs and at the response-end for LDs for security and permissions respectively.
            Tiny buffers are maintained to ensure all LD/ST are committed in-order.

            We also maintain the confidentiality of local pages.
            A trusted process' local memory addresses will have the A-bits tagged.
            We add a memory encryption engine for encrypting these pages using the host's private key ($K_{host}$).
            The design therefore provides both confidentiality and integrity for a trusted process's local data.


        \subsubsection{Extensions to the Fabric Manager (FM)}
            The FM serves as the global root-of-trust for \papertitle.
            Today, the only role of the FM is to set up the hardware tables for access control (\S\ref{sec:background}).
            We extend its role to manage the permission table in the SDM, and, generate public labels for process authentication for each host.

            Upon granting access to a process on a host, the FM issues a public label ($L_{exp}$) that cryptographically binds a \id (\context) to its permitted memory ranges.
            The label is stored in the label section of the metadata and the entry, which has a defined range, is committed to the permission table.
            In addition, the FM is also responsible for optimizing the permission table to minimize the total number of entries.

\section{Discussion}
    \label{sec:discussion}
    The \papertitle's identity primitive serves as a universal architectural building block.
    By anchoring identity in hardware, the abstraction scales independent of the host OS.
    While this work evaluates a process-level implementation, the following sections validate the security properties inherent to the primitive and justify our core design choices.

    \subsection{Security Analysis}
        \label{sec:security}
        Our claim is this paper is that we guarantee the \textit{confidentiality} of trusted process' data and trust the disaggregation hardware for the \textit{availability}.
        \subsubsection{Malicious host processes}
            The host can spawn multiple processes and even cause a denial-of-service attack, which tries to access the SDM over the fabric.
            \papertitle validates the trusted \id via the SPACE.
            The \id cannot be forged by any other process.
            Any LD/ST generated by a trusted process is further checked by the permission checker.
            The checker raises a fault locally when a non-trusted process (without the tagged \textit{A-bits}) tries to access the remote memory region directly.
            
        \subsubsection{Compromised system software}
            We assume a strong threat model in which the OS may collude with untrusted processes to violate the confidentiality of a trusted process' local and remote data.
            We therefore use hardware-defined identity primitive (\id).
            This prevents the OS from forging context labels.

            Although the OS can still modify page tables, such manipulation does not compromise remote memory.
            If the kernel maps the SDM region into an untrusted process's address space, SPACE rejects the access as authentication fails, and the permission checker blocks any unauthorized LD/ST.

            A stronger threat arises when the kernel attempts to compromise the local memory \textit{confidentiality} of a trusted process.
            Since local memory is untrusted in our model, the kernel may remap the page tables of a trusted process and a malicious process to the same physical memory region.
            The trusted process remains unaware of the remapping and continues to use its local pages, while the malicious process gains read access to those same pages, violating \textit{confidentiality}.

            To address such threats, \papertitle employs a memory encryption engine that encrypts all local memory pages belonging to trusted processes, identified by their tagged \textit{A-bits}.
            This ensures \textit{confidentiality} even if the OS aliases page tables.
            While encrypted data prevents unauthorized reads, a malicious process could still overwrite ciphertext, affecting \textit{integrity}.
            When such writes are detected in the software, \papertitle identifies the presence of compromised system software, satisfying our target threat model from \S\ref{sec:threat_model}.

        \subsubsection{Malicious DMA devices}
            \papertitle assumes that SDM is accessed primarily via CPU LD/ST operations using \texttt{/dev/dax}~\cite{famfs}.
            Unlike system DRAM, a DAX region is not part of the system memory map.
            It is exposed as a character device and mapped into a processes' virtual address space.
            Consequently, DMA-capable devices, such as GPUs and other accelerators, cannot directly access this region without explicit peer-to-peer mappings.
            This distinction reinforces the gap we target, \textit{i.e.}, existing PASID-based Shared Virtual Addressing (SVA) mechanisms only apply to device-initiated DMA and do not cover CPU-originated accesses to SDM.
            Therefore, hardware enforcement at the CPU egress point remains essential for guaranteeing process-level isolation in scenarios where the OS cannot be trusted.
    
    \subsection{Design Choices and Limitations}
    
        \papertitle introduces additional wiring overhead by tagging PA with the \textit{A-bits}.
        The design is inspired by the C-Bit of AMDSEV~\cite{amdsev}.
        We justify our design choice of extending the PA as today's \textit{state-of-the-art} architectures uses up to 57 out of 64 bits~\cite{intel-5-level,riscv-sv57}.
        This allows flexibility of using the remaining unused bits to tag the address space.
        Intel hardware already adds support to extend the upper 7 bits of the physical addresses in the kernel using Linear Address Masking~\cite{lam-kernel}.
        We allow up to 127 processes per host to ensure 64 bits address compatibility.
        For simplicity, we choose to place the HWPID tag bits into the most significant bits of the address, which does not cause conflicts with the existing wiring for the TLB, the caches and ensures CXL compatibility. 


        An alternate method of protecting memory was via a reverse mapping table ($PA \rightarrow (VA, HWPID)$) per host, similar to AMDSEV~\cite{amdsev}.
        Such tables are also popular to identify hot cache lines for page migration in CXL-related prior works~\cite{hopp}.
        A reverse mapping from PA to (VA, PID) allows the hardware to infer which process owns a given physical page without modifying the address format.
        However, given the current programming model for SDM, the base pointer to the \texttt{mmap} region is fixed and addresses are accessed based on offsets.
        We chose to adopt a simple permission table where the mapping from PA to (host, (\context)) is enough to ensure that the OS cannot modify the page table to remap the SDM.

        Based on our threat model, our design guarantees \textit{confidentiality} to the data at all times and \textit{integrity} to the data in the SDM.
        However, we rely on the CXL hardware to function correctly and provide \textit{availability}.
        This distinguishes \papertitle from a TEE.
        If the trusted process is compromised at runtime, \papertitle fails both \textit{confidentiality} and \textit{integrity} guarantees.

\section{Evaluation}
    \label{sec:eval}

    \begin{table}[t]
        \centering
        \resizebox{0.4\textwidth}{!}{%
        \begin{tabular}{c|c|}
            \hline
            \rowcolor[HTML]{C0C0C0} 
            \textbf{Parameter} & \textbf{Specification} \\ \hline \hline
            Number of hosts & up to 8 (incl. FM) \\
            Number of devices & 1 \\ \hline
            CPU type & TimingSimpleCPU \\
            CPU core count & 8 (1 for the FM) \\
            Frequency & 4 GHz \\ \hline
            L1 cache (per core) & 32 KiB + 32 KiB (I + D) \\
            L2 cache (per core) & 1 MiB \\                
            L3 cache (shared) & 16 MiB \\ \hline                 
            Memory technology & DDR4 DRAM \\
            Memory frequency & 2400 MHz \\
            Number of local memory channels & 2 \\
            Number of remote memory channels & 4 \\
            Local memory size & 16 GB \\                 
            Peak local memory bandwidth & 38.4 GiB/s \\
            Remote memory size & 16 GiB \\
            Peak remote memory bandwidth & 76.8 GiB/s \\ \hline
            Operating System & Ubuntu 22.04.4 (Jammy Jellyfish) \\
            Kernel version & 6.9.9 \\ \hline \hline
            \end{tabular}%
        }
        \caption{Simulated system specifications.}
        \label{tab:system-params}
    \end{table}

    \subsection{Setup and Workloads}
        \label{subsec:workloads}
        Our goal is to understand the performance implications of \papertitle.
        We used a modified version of GAPBS~\cite{gapbs} to share a graph across several hosts.
        We have used CXL-ClusterSim~\cite{cxl-clustersim} for evaluating \papertitle.
        CXL-Clustersim uses gem5~\cite{gem5} to model the hosts for fidelity and SST~\cite{sst} to model the shared disaggregated memory and parallelize the simulation~\cite{gem5sst}.
        \papertitle hardware is implemented in gem5.
        The modeling goal is to have a realistic CXL.mem implementation.
        The CXL device, modeled in SST, is implemented as \texttt{timingDRAM} modeled as four-channel 2400 MHz DDR4 DRAM.
        The latencies of translation and lookup of remote memory addresses is implemented at the gem5-SST bridge~\cite{gem5sst}, that connects the two simulators together. 
        The bridge is modified to support gem5-style checkpointing and KVM for fast-forwarding.
        The timing parameters of the CXL protocol is based on prior characterization and research~\cite{pond,dassharmax,cxl-main,mem-disaggregation}.
        
        There are eight hosts. 
        For simplicity, instead of modeling CXL.io for FM, we use shared CXL.mem for host and FM communication.
        Hosts still interact with the permission table using UIO~\cite{uio-kernel}.
        Host 0 is responsible for allocating the data for the workloads.
        For GAPBS, the rest of the hosts (1--6) run a specific kernel.

    \subsection{Experiments}
        \label{subsec:exp}
        To evaluate \papertitle, we compare the performance slowdown due to permission checks at each LD/ST against a baseline CXL 3.0 system (\textit{cxl}).
        We show performance in terms of cycles per instruction (CPI) of the graph kernel's \textit{region of interest}.
        The baseline is a system described in Table~\ref{tab:system-params} without any permission checks.
        On a real CXL system, once a VA is translated to a PA, a CXL packet is created for the PA and sent to the CXL network until the memory device is reached.
        In our \textit{cxl} system, we model the address translation correctly to establish a performance baseline for us to compare against \papertitle.
        We simulate the first 4 billion cycles in the \textit{region of interest} for each workload running on each host.
        
        Additional latencies are modeled for \textit{A-bit} tag comparison and memory encryption.
        We use a hardware-efficient encryption engine implementation similar to prior works~\cite{amdsev,intel-sgx}, where encryption per cache-line is at most 1 cycle.
        We repeat the experiments for sensitivity analysis with the permission cache.
        In addition to evaluating \papertitle, we also implement several related works and compare the overall performance implications for enforcing permissions.
        
        






\section{Results}
    \label{sec:results}
    In this section, we discuss our results based on the evaluation strategy listed in \S\ref{sec:eval}.
    \label{sec:perf} This section is divided into three primary subsections that discuss the 
    \textit{(a)} performance,
    \textit{(b)} overheads, and,
    \textit{(c)} compares \papertitle with existing works.

    \subsection{Performance Analysis}
        While \id verification via SPACE incurs an insignificant one-time cost, we need to analyze the permission checker in details to understand the actual permission overhead.

        \subsubsection{Baseline and Scalability}
            \label{sec:scaling}
            This experiment establishes the \textit{cxl} baseline, modeled after CXL 3.0, and quantifies \papertitle's end-to-end overhead in the best-case sharing layout, when a single permission entry (\textit{space-control-1e}) spanning the SDM region.
            This setting isolates the intrinsic cost of per-access authentication and permission checking from metadata fragmentation effects, and it lets us evaluate how the egress checker scales as more hosts participate.
            For each local access we verify \id, check for the \textit{A-bits} on the PA, and, encrypt the pages.
            For each SDM access we consult the permission table for the current \id.

            We run GAPBS kernels while scaling the number of hosts accessing the same SDM region from 1 to 8 in powers of two.
            We use four representative kernels: \texttt{pr}, \texttt{bfs}, \texttt{bc}, \texttt{tc}; ordered from more regular to more random access patterns, and report CPI normalized to the \textit{cxl} baseline (no checks).
            The single-entry configuration keeps the permission lookup path constant, so any change in CPI reflects the rate of permission lookups rather than longer lookup latency.

            \begin{figure}[t]
                \centering
                \mbox{
                \subfigure[Performance of the system with a single permission entry (best-case layout). CPI is normalized to the \textit{cxl} baseline]
                    {
                    \label{fig:scaling-1e}
                    \includegraphics[width=0.23\textwidth]{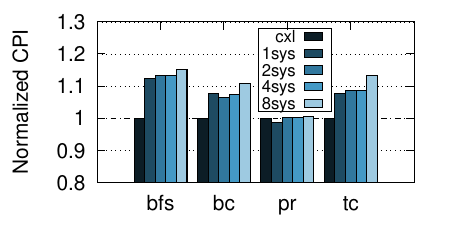}
                    }
                \subfigure[CPI of all GAPBS kernels running concurrently on each host, normalized to the eight host \textit{cxl} baseline.]
                    {
                    \label{fig:all-cpi-1e}
                    \includegraphics[width=0.23\textwidth]{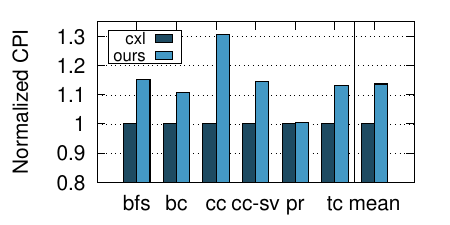}
                    }
                }
                
                \caption{Reported CPI in \papertitle normalized to \textit{cxl}.}
                \label{fig:scaling-all}
            \end{figure}

            Figure~\ref{fig:scaling-1e} shows the CPI for the four workloads when the number of systems are scaled.
            \papertitle imposes a small overhead at one host (7.3\% on average than \textit{cxl}) that grows sub-linearly to 12.1\% at eight hosts.
            This trend demonstrates that the permission checker scales without becoming a bottleneck, \textit{i.e.} adding hosts increases the number of permission lookups, but the pipeline and caches keep lookup latency bounded.
            The workload ordering matches intuition: \texttt{pr} stays closest to baseline due to streaming locality, and, the least LLC misses; whereas the rest of the workloads incur more lookups and pay a larger penalty.
            This reinforces the fact that locality and LLC misses drives overhead.

            In the eight-host multiprogrammed setup of Figure~\ref{fig:all-cpi-1e}, the same story persists, \textit{i.e.} \texttt{pr} remains near baseline (0.6\%), while \texttt{cc} sits higher (23.4\%) because \texttt{cc} has a higher LLC miss rate which generates more irregular requests.
            The key takeaway for deployment is that when tenants map few, contiguous SDM ranges, \papertitle's cost is predictable even as tenancy scales.
            The next subsection stress-tests the opposite extreme: \textit{worst-case} fragmentation to bound overhead when the permission table becomes deep and per-access lookups dominate.

        \subsubsection{Sensitivity to Permission Table Fragmentation}
            \label{sec:wc-results}
            Having established the single-entry baseline, we now stress the design with \textit{worst-case} fragmentation to expose the cost of the sorted permission table when every access requires a deeper lookup.
            This experiment isolates the central trade-off as we replace the storage overhead of flat tables with lower metadata but higher lookup time.
            We then investigate the lookup overhead gets when the permission table is maximally deep.
            Concretely, we compare the best case (a single permission entry spanning the SDM, \textit{space-control-1e}) to the worst case where every 4 KiB range has its own distinct entry (\textit{wc}), so each SDM reference can trigger a binary search in a large table.
            
            We repeat the baseline experiment under this worst-case layout and report normalized CPI alongside two indicators of lookup pressure: the binary-search occupancy (how many table probes per lookup) and the permission lookups per kilo-instructions (PLPKI).
            This framing makes the mechanism's behavior transparent as
            \textit{(a)} CPI reflects end-to-end impact,
            \textit{(b)} occupancy shows how deep the table is exercised, and,
            \textit{(c)} PLPKI captures how often the checker is invoked by the workload.
            
            \begin{figure}[t]
                \centering
                \mbox{
                \subfigure[Reported CPI normalized to the baseline when the number of hosts sharing the SDM were increased in powers of two.]
                    {
                    \label{fig:scaling-wc}
                    \includegraphics[width=0.23\textwidth]{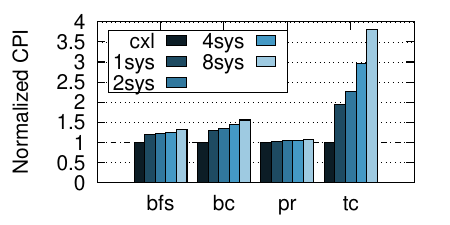}
                    }
                \subfigure[Reported PLPKI for a single permission entry case (\textit{1e}) and a complete table fragmentation case (\textit{wc}).]
                    {
                    \label{fig:plkpi}
                    \includegraphics[width=0.23\textwidth]{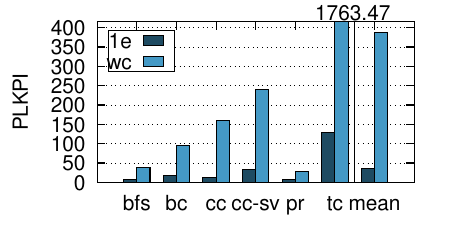}
                    }
                }
                \caption{Reported CPI in \papertitle normalized to \textit{cxl} when the memory is completely fragmented.}
                \label{fig:scaling-all-wc}
            \end{figure}
\begin{figure}[!b]
                \centering
                \includegraphics[width=1\columnwidth]{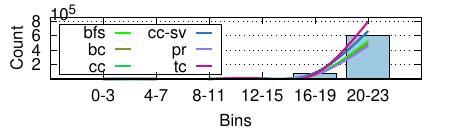}
                \caption{The PDF of binary searches for a unique permission entry. The bars represent the average histogram sorted by the bins.}
                \label{fig:histograms-occupancy}
            \end{figure}

            Figure~\ref{fig:scaling-wc} shows that fragmentation amplifies overhead in proportion to access irregularity: \texttt{tc}, which intentionally skips relabeling and exhibits poor locality, rises the most (3.8$\times$ \textit{vs. cxl}), whereas \texttt{pr} remains close to baseline (5.7\%).
            Figure~\ref{fig:histograms-occupancy} corroborates the mechanism as \texttt{tc} drives the highest binary-search occupancy, and Figure~\ref{fig:plkpi} confirms that PLPKI tracks this increase as more distinct, scattered references mean more frequent, deeper permission lookups.
            By contrast, \texttt{bfs} and \texttt{bc} sit between \text{pr} and \texttt{tc}, consistent with their frontier-driven irregularity but better reuse than \texttt{tc}.
            
            The takeaway is that \papertitle's overhead under fragmentation is lookup-dominated and workload-dependent.
            That is exactly the cost we chose in exchange for low storage overhead and OS-independent process isolation.
            Importantly, the worst case is both diagnosable (high occupancy, high PLPKI) and actionable: operators can coalesce ranges or align shared allocations to cut table depth, and \S\ref{sec:permission-cache} shows that even a small permission cache amortizes lookups and restores performance.
            Together, these results bound the upper cost of fine-grained permissions and show how to steer deployments toward the single-entry or low-fragmentation regime, where \papertitle remains predictable and modest.

        \subsubsection{Memory-Access Split and Memory Bandwidth}
            \label{sec:memory}

            \begin{figure}[t]
                \centering
                \mbox{
                \subfigure[Stacked graphs showing the count of data traffic (darker shade) and permission traffic (lighter shade).]
                    {
                    \label{fig:memory-split}
                    \includegraphics[width=0.27\textwidth]{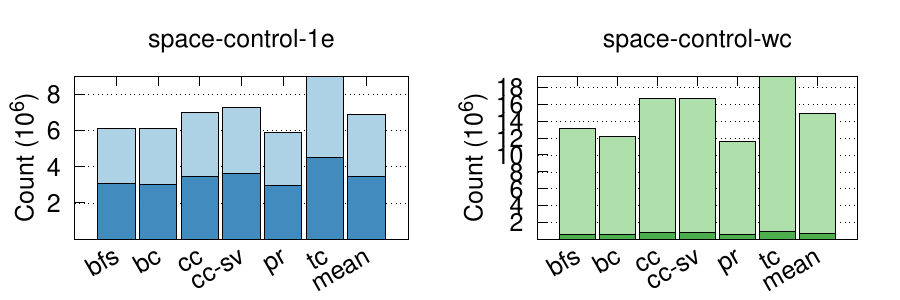}
                    }
                \subfigure[Remote memory bandwidth usage reported by each host.]
                    {
                    \label{fig:memory-bw}
                    \includegraphics[width=0.18\textwidth]{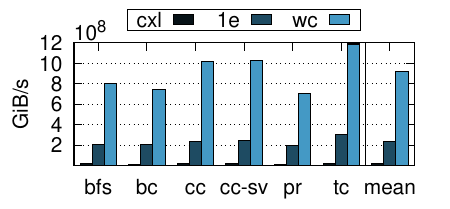}
                    }
                }
                \caption{Memory traffic analysis of \papertitle.}
                \label{fig:memory-all}
            \end{figure}

            We quantify how much fabric activity comes from permission packets versus data packets, because both contend for the same CXL links and device queues.
            Figure~\ref{fig:memory-split} shows that, under fragmented tables, irregular kernels (\texttt{tc}, \texttt{bc}) drive a large share of permission packets.
            Figure~\ref{fig:memory-bw} reports per-host bandwidth of the remote memory.
            As the permission share falls (\textit{1e vs. wc}), hosts experience fewer permission-induced stalls and sustain higher effective data throughput, while the device spends proportionally more cycles servicing data than probing the table.
            
            \papertitle's cost is governed by lookup frequency, not bulky control transfers.
            Two practical levers, \textit{i.e.} keep shared ranges coarse and provision a small permission cache, can restore a fabric profile where permission traffic is a minor control-plane slice and data bandwidth tracks application demand.

        \subsubsection{Breakdown of Performance Contributors}
            \label{sec:breakdown}
            \begin{figure}[!b]
                \centering
                \mbox{
                \subfigure[Stall latency across all the GAPBS workloads.]
                    {
                    \label{fig:stall-all}
                    \includegraphics[width=0.23\textwidth]{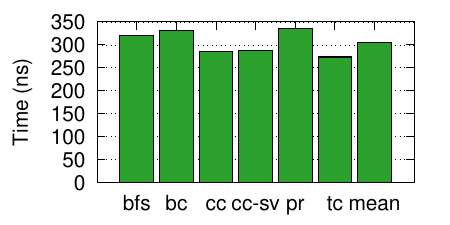}
                    }
                \subfigure[Stacked latencies of the three factors contributing towards performance slowdown.]
                    {
                    \label{fig:stacked-time}
                    \includegraphics[width=0.23\textwidth]{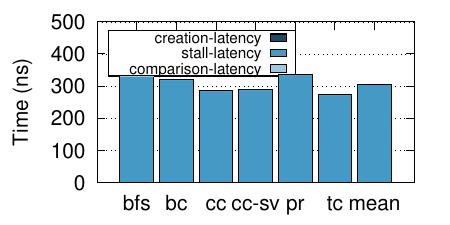}
                    }
                }
                \caption{Performance breakdown analysis.}
                \label{fig:performance-all}
            \end{figure}
            To understand the source of the overhead, we instrument the permission checker pipeline and attribute runtime to three components: \textit{(a)} permission request creation, \textit{(b)} permission lookup latency, and \textit{(c)} enforcement latency, \textit{i.e.} the time a LD/ST response is stalled until all corresponding permission entries arrive.
            Creation is largely depended on the circuit design and is small; the other two depend on the table organization and on how many lookups each workload triggers, tying this analysis directly to \S\ref{sec:wc-results}'s fragmentation results.

            Figure~\ref{fig:stacked-time} shows that enforcement stalls dominate the slowdown (99.95\% in our configuration), while A-bit comparisons are negligible (0.003\%).
            Intuitively, irregular workloads and fragmented tables inflate both lookup depth and lookup rate, which lengthens stalls because more responses must be matched with more (and deeper) permission entries before commit.

        \subsubsection{Performance Enforcement Latency}
            \label{sec:stall}

            \S\ref{sec:breakdown} shows enforcing permissions impacts the response path the most.
            We issue LD/ST and permission requests out-of-order, but a LD/ST response cannot commit until all of its corresponding permission entries have arrived.
            We call the resulting wait time as the enforcement or the stalling latency.
            Figure~\ref{fig:stall-all} reports the average tail latency across workloads.
            It reflects the time the pipeline spends buffering responses to preserve in-order architectural commit while \papertitle verifies access rights.
            
            \begin{figure}[t]
                \centering
                \includegraphics[width=1\columnwidth]{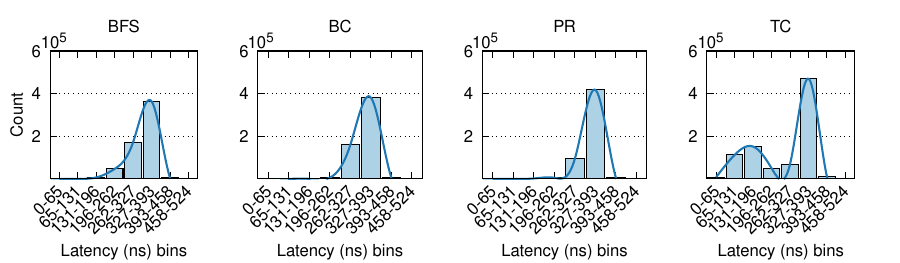}
                \caption{Histogram of permission enforcement latency.
                        The line represents the PDF of the latency distribution.}
                \label{fig:stall-histograms}
            \end{figure}

            Consistent with \S\ref{sec:wc-results}, more random access patterns yield longer stalls because they trigger more and deeper permission lookups.
            The histograms in Figure~\ref{fig:stall-histograms} make this visible, where kernels like \texttt{tc} and \texttt{bc} exhibit heavier stalls than \texttt{pr}.
            The takeaway is that enforcement latency is a lookup-dominated cost; reducing lookup frequency (fewer, coarser ranges) or amortizing lookups (a small permission cache) directly shortens the stall time.
            This sets up \S\ref{sec:permission-cache}, where we show that modest caching materially reduces enforcement time and restores end-to-end responsiveness.

        \subsubsection{Performance Cache Efficacy}
            \label{sec:permission-cache}
            \begin{figure}[!b]
                \centering
                \mbox{
                \subfigure[Reported permission cache miss ratio with increasing permission cache size in powers of two.]
                    {
                    \label{fig:cache-miss-ratio}
                    \includegraphics[width=0.18\textwidth]{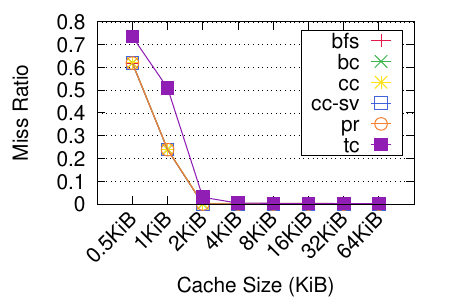}
                    }
                \subfigure[Reported CPI of the system with increasing number of cache sizes. Numbers are normalized to a \papertitle system with complete memory fragmentation without caching.]
                    {
                    \label{fig:cache-wc}
                    \includegraphics[width=0.28\textwidth]{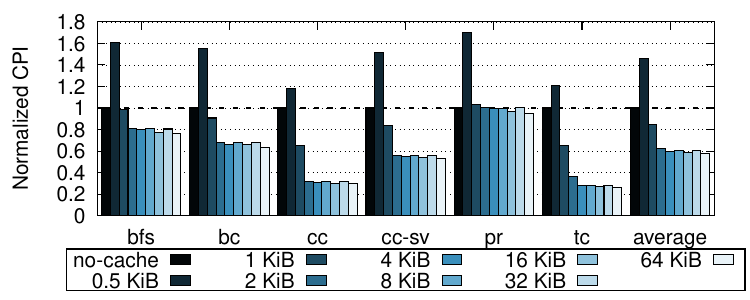}
                    }
                }
                \caption{Measured performance efficacy of \papertitle with a permission cache. All systems are assumed to be completely fragmented, as mentioned in \S\ref{sec:wc-results}.}
                \label{fig:caching-all}
            \end{figure}
            The previous sections showed that \papertitle's overhead is lookup-driven (\S\ref{sec:wc-results}) and expresses as enforcement stalls (\S\ref{sec:stall} and \S\ref{sec:breakdown}).
            A small permission cache amortizes repeated lookups as it reduces PLPKI and binary-search depth, which shortens enforcement stall latency and restores throughput (\S\ref{sec:memory}).

            We sweep cache sizes from 0.5 KiB (8 entries) to 64 KiB (1024 entries) and report miss ratio (Figure~\ref{fig:cache-miss-ratio}) and CPI normalized to the uncached worst-case fragmentation (Figure~\ref{fig:cache-wc}).
            In this stress setting, where there are distinct entry per 4 KiB range, the cache targets the mechanism's worst behavior (many deep lookups).
            
            There is a pronounced elbow in the cache-benefit curve: a modest permission cache delivers most of the gain.
            With just 2 KiB of cache, the hit rate reaches 99.9\% and the CPI overhead improves by 2.3$\times$ over the uncached \papertitle configuration on GAPBS.
            The sharp drop in miss ratio from 1 KiB to 2 KiB stems from the way the sorted permission table is searched: a binary search repeatedly revisits a small set of internal decision nodes across lookups, while the leaf (the final permission entry for the specific range) is the part most likely to be uncached.
            
            For a 16 GiB shared memory, the binary search touches at most 24 nodes at worst.
            A 2 KiB cache holds 32 entries, which exceeds the depth and keeps most internal nodes resident.
            Only the leaf entries tends to miss.

            This scaling intuition generalizes.
            For 1 TiB shared memory (up to $lg(2^{40}) = 40$ probes), 4 KiB cache (64 entries) exceeds the worst-case search depth, amortizing the lookup cost.
            This makes cache auto-tuning straightforward, \textit{i.e.} choose a capacity that meets or slightly exceeds $lg(\text{table size})$. 
            
            Compared to the CXL baseline, a 16 KiB cache leaves only 3.3\% marginal performance overhead.
            Consistent with our previous results, caching also trims enforcement stalls, and benefits show diminishing returns beyond 4 KiB in the worst-case fragmentation for most workloads.
            A small permission cache converts \papertitle from lookup-bound and stall-heavy to data-dominant, aligning with the design flow we established so far.

        \subsubsection{Permission Revocation}
            Updating permissions follows the CXL's back invalidate snoop (\texttt{BISnp}) protocol~\cite{cxl-3.1-spec}.
            When a host updates the permission table via a new entry and the entry is approved and pushed into the table by the FM, a CXL \texttt{BISnp} is sent on the network.
            Permission caches on hosts caching that entry will automatically invalidate the permission entry from its own cache.
            The latency of performing a permission revocation is the same as CXL's \texttt{BISnp} latency.

            One limitation of permission revocation is that it fails to prevent on-flight instructions getting committed.
            There exists a tiny window of at most two \texttt{BISnp} where an attacker can violate isolation.
            This problem however is the CXL's cache coherence problem~\cite{cxl-3.1-spec,hdcs-paper}, which is orthogonal to our work.

    \subsection{Hardware Overhead}
        In \papertitle, a memory range of an entry can be of any arbitrary length.
        For practicality and future-proofing, we limit the minimum length to be 4 KiB, the same as the smallest unit of an OS-level page size.
        This implies that in the worst case where every 4 KiB page of the SDM will have a 64 Byte entry, resulting in a maximum of 1.56\% overhead.

        We use 32 permission miss status holding registers to keep track of all the permission entry misses.
        The response buffer for permission enforcement holds up to 32 responses.
        This totals 4 KiB of additional SRAM storage.

        The cryptographic engine follows a similar implementation to prior works~\cite{hwimpl}.
        We add two 64-bit shadow registers \texttt{label} and \texttt{authentication result} at the CPU.
        SPACE uses minimal logic and wiring overhead for comparing \id and generating labels.
        The total SRAM storage needed to maintain the key (64 bits), counter (64 bits), free HWPID list (128 entries) and public label, \textit{i.e.} $L_{exp}$ (64 bits) is 1.048 KiB.
        The permission checker uses five comparators to generate permission requests and enforce permissions.

    \subsection{Comparison to Prior Mechanisms}
        \label{sec:comparison}

\begin{figure}[t]
                \centering
                \includegraphics[width=1\columnwidth]{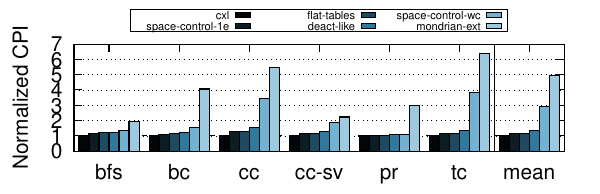}
                \caption{CPI of \papertitle and several other related works. The CPI is normalized to the baseline \textit{cxl}. None of the systems have any performance enhancing permission caches for a fair comparison.}
                \label{fig:comparison}
            \end{figure}

        In addition to the \textit{cxl} baseline, we also compare the performance slowdown to an ideal \textit{flat-table} structure similar to Border-Control~\cite{bordercontrol}, and, other related works: \textit{deact-like}~\cite{deact} and \textit{mondrian-ext}~\cite{mmp}.
        Figure~\ref{fig:comparison} shows the CPI normalized to the baseline across all the other techniques.
        The \textit{flat-table} structure requires at most one permission table entry lookup per remote memory access.
        Compared to the baseline, the slowdown is 13.1\%, which is comparable to \textit{space-control-1e}.
        \textit{space-control-1e} slightly performs better as there is a single entry at a fixed location unlike \textit{flat-tables}, where each PPN reads its corresponding permissions from a different region of the SDM. 
        The total storage on the other hand, is given by:
        \begin{equation}
            \begin{split}
            N_{hosts} \times M_{processes} \times P_{pages} \times ENTRY_{size} \\
            = 256 \times 128 \times (16 \text{GiB} / 4 \text{KiB}) \times (2b / 8) = 32 \text{GiB}
            \end{split}
        \end{equation}
        The storage overhead is 200\% than the total remote memory size and $128.2\times$ larger than \papertitle.

        \textit{deact-like} is our implementation of DeACT~\cite{deact}.
        We maintain an address-permission mapping per host per 4 KiB SDM page instead of 1 GiB page for a fair comparison.
        We adopt the table for sharing 16 GiB of memory across 256 hosts, where each host can have up to 128 processes.
        The total storage for sharing one process is given as:
        \begin{equation}
            SIZE_{mapping\_table} + SIZE_{sharing\_bitmap} = 0.156 \text{GiB}
        \end{equation}
        which is only 0.9\% total overhead.
        However, DeACT's threat model does not trust the OS but lets the OS manage the page tables for all the processes sharing SDM. 
        If we scale to 128 processes outside the host OS' control, the replication of the mapping table and the bitmap takes 20 GiB of metadata, which is 125\% the size of the total remote memory size, and $80.1\times$ larger than \papertitle.
        The average CPI of \textit{deact-like} is 32.66\% higher than \textit{space-control-1e} as there are at most two permission entry lookups in DeACT: the owner mapping entry and the shared bitmap.

        We also extend Mondrian to support memory disaggregation (\textit{mondrian-ext}).
        The original Mondrian solution does not distinguish between remote or local addresses as mondrian permission tables are set up for every process.
        We maintain at least two domains: \textit{(a)} local memory processes and \textit{(b)} the real workload, which spawns across both the local and the remote memory.

        Domains in Mondrian and permission entries in \papertitle are user-defined.
        For the workloads we have, one domain or entry in \papertitle is enough to show the principle of least privilege is satisfied in these works.
        However, for complete fairness, we compare a single entry as the best case (\textit{1e}), and, every 4 KiB entry as the worst case scenario (\textit{wc}) against \textit{cxl}, \textit{flat-table} and \textit{deact}.
        Mondrian maintains two protection domains: \textit{(a)} for the SDM region, and, \textit{(b)} single permission entry per host for the local memory.

        In terms of storage overhead, both Mondrian and \papertitle are similar.
        Mondrian's sorted segment table entries are smaller however, they need to replicate their permission table (sorted segment table) per host, linearly increasing the total storage overhead.
        \papertitle performs 4.3$\times$ better than \textit{mondrian-ext} in terms of performance as the scope of \papertitle is only for shared memory accesses.
        In other words, \papertitle extends the principles of Mondrian for an untrusted OS in the era of CXL and shared disaggregated memory.



\section{Related Works}
    \label{sec:related}

    \paragraph{Memory disaggregation.}
    \S\ref{sec:background} gives a brief introduction to disaggregated memory.
    Prior works like DeACT~\cite{deact} relies on Gen-Z's host-side zMMU~\cite{genz} and two-stage translation to decouple address translation from permission checks.
    CXL removes the host zMMU and reuses the CPU's TLB~\cite{cxl-3.1-spec}, so \papertitle enforces authorization after the last-level cache.
    RDMA-based works like~\cite{finemem} is orthogonal to CXL.

    
    
    \paragraph{OS-managed mechanisms.}
    OS-managed protections via \texttt{namespaces} and \texttt{cgroups}~\cite{cgroups}, MPK and ARM MTE provide intra-host protection but assume a trusted kernel.
    LegoOS scales disaggregation but also trusts the OS for isolation.
    \papertitle removes the OS from the enforcement path (\S3).~\cite{intel-mpk,arm-mte,legoos}
    Border Control~\cite{bordercontrol} sandboxes accelerators with page-level checks under a trusted OS.
    \papertitle is inspired by this work and builds on this insight by treating untrusted hosts analogous to untrusted accelerators, enforcing access control through a hardware permission checker integrated into the memory access path.

    \paragraph{Domain/capability designs.}
    Mondrian~\cite{mmp,mondrix} introduces the concept of domains and uses a sorted protection table to trade-off storage for higher lookup latency. 
    \papertitle adopts the sorted-table idea but enforces at the CPU egress with process-level isolation and an untrusted OS (\S\ref{sec:method}).
    CHERI-style capability systems offer principled fine-grained authority but require pervasive ISA/compiler/OS changes~\cite{cheri-paper,cheri-flexible,cheri-sp}.
    Bresniker \textit{et al.}~\cite{cheri-rackscale} proposed the addition of memory-side capability check to extend CHERI to the disaggregated memory space.
    The memory-side capability check is for address range check per process per host, creating a high metadata overhead and restricting scalability.
    Furthermore, it requires a trusted OS for process coordination. 
    We target deployability on existing CPUs and CXL fabrics using simple techniques like \texttt{LD\_PRELOAD}.
    
    \paragraph{TEEs and enclave models.}
    TEEs~\cite{intel-sgx,intel-tdx,amdsev,elasticlave} provide isolation and attestation but are ill-suited for multi-host shared ranges~\cite{cxl-3.1-spec,cxl-4-spec} and add non-trivial overhead~\cite{ayaz-ipdps}.
    Sharing memory enclaves on a single host is an active research topic~\cite{elasticlave,data-enclaves}, however commercially available TEEs are yet to adopt it.
    Minerva~\cite{minerva} is an extension to the TEE design for disaggregated memory, where the authors made a fault-tolerant metadata management scheme for disaggregated memory.
    ShieldCXL~\cite{taco-oram-cxl} proposes an oblivious TEE for CXL systems.
    Li~\textit{et al.}~\cite{micro-cxl} proposes an efficient technique to encrypt CXL flits for designing a TEE on a CXL system.
    \papertitle enforces authorization without enclaves.
    
\section{Conclusion} 
    \label{sec:conclusion}
    
    This work addresses a critical gap in disaggregated memory systems: the lack of identity primitive on CXL fabrics.
    \papertitle provides a scalable hardware-software co-design abstraction that enforces process-level isolation without relying on the OS.
    The design remains fully compatible with existing virtual memory abstractions.
    Our design uses a space-efficient metadata structure and leverages the CXL fabric manager to scale across many hosts and processes, incurring only 1.56\% memory overhead and 3.3\% performance overhead.
    By reconciling strong security with system scalability, \papertitle establishes a practical foundation for secure disaggregated memory.
    For future work, we plan on adding identity primitive to virtual machines and enclaves for sharing memory across hosts, and providing \textit{availability} at all times.

\bibliographystyle{plain}
\bibliography{references}

@misc{cxl_sharing_2024,
      title={Memory Sharing with CXL: Hardware and Software Design Approaches}, 
      author={Sunita Jain and Nagaradhesh Yeleswarapu and Hasan Al Maruf and Rita Gupta},
      year={2024},
      eprint={2404.03245},
      archivePrefix={arXiv},
      primaryClass={cs.ET},
      url={https://arxiv.org/abs/2404.03245}, 
}

@inproceedings{mmp,
author = {Witchel, Emmett and Cates, Josh and Asanovi\'{c}, Krste},
title = {Mondrian memory protection},
year = {2002},
isbn = {1581135742},
publisher = {Association for Computing Machinery},
address = {New York, NY, USA},
url = {https://doi.org/10.1145/605397.605429},
doi = {10.1145/605397.605429},
booktitle = {Proceedings of the 10th International Conference on Architectural Support for Programming Languages and Operating Systems},
pages = {304–316},
numpages = {13},
location = {San Jose, California},
series = {ASPLOS X}
}

@misc{opencapi,
    author = "{IBM}",
    title = "{OpenCAPI (Open Coherent Accelerator Processor Interface)}",
    year = 2014,
    url = {https://docs.kernel.org/userspace-api/accelerators/ocxl.html}
}

@misc{gem5,
      title={The gem5 Simulator: Version 20.0+}, 
      author={Jason Lowe-Power and Abdul Mutaal Ahmad and Ayaz Akram and Mohammad Alian and Rico Amslinger and Matteo Andreozzi and Adrià Armejach and Nils Asmussen and Brad Beckmann and Srikant Bharadwaj and Gabe Black and Gedare Bloom and Bobby R. Bruce and Daniel Rodrigues Carvalho and Jeronimo Castrillon and Lizhong Chen and Nicolas Derumigny and Stephan Diestelhorst and Wendy Elsasser and Carlos Escuin and Marjan Fariborz and Amin Farmahini-Farahani and Pouya Fotouhi and Ryan Gambord and Jayneel Gandhi and Dibakar Gope and Thomas Grass and Anthony Gutierrez and Bagus Hanindhito and Andreas Hansson and Swapnil Haria and Austin Harris and Timothy Hayes and Adrian Herrera and Matthew Horsnell and Syed Ali Raza Jafri and Radhika Jagtap and Hanhwi Jang and Reiley Jeyapaul and Timothy M. Jones and Matthias Jung and Subash Kannoth and Hamidreza Khaleghzadeh and Yuetsu Kodama and Tushar Krishna and Tommaso Marinelli and Christian Menard and Andrea Mondelli and Miquel Moreto and Tiago Mück and Omar Naji and Krishnendra Nathella and Hoa Nguyen and Nikos Nikoleris and Lena E. Olson and Marc Orr and Binh Pham and Pablo Prieto and Trivikram Reddy and Alec Roelke and Mahyar Samani and Andreas Sandberg and Javier Setoain and Boris Shingarov and Matthew D. Sinclair and Tuan Ta and Rahul Thakur and Giacomo Travaglini and Michael Upton and Nilay Vaish and Ilias Vougioukas and William Wang and Zhengrong Wang and Norbert Wehn and Christian Weis and David A. Wood and Hongil Yoon and Éder F. Zulian},
      year={2020},
      eprint={2007.03152},
      archivePrefix={arXiv},
      primaryClass={cs.AR}
}

@article{sst,
author = {Rodrigues, A. F. and Hemmert, K. S. and Barrett, B. W. and Kersey, C. and Oldfield, R. and Weston, M. and Risen, R. and Cook, J. and Rosenfeld, P. and Cooper-Balis, E. and Jacob, B.},
title = {The structural simulation toolkit},
year = {2011},
issue_date = {March 2011},
publisher = {Association for Computing Machinery},
address = {New York, NY, USA},
volume = {38},
number = {4},
issn = {0163-5999},
url = {https://doi.org/10.1145/1964218.1964225},
doi = {10.1145/1964218.1964225},
journal = {SIGMETRICS Perform. Eval. Rev.},
month = {mar},
pages = {37–42},
numpages = {6}
}

@inproceedings{gem5sst,
author = {Hsieh, Mingyu and Pedretti, Kevin and Meng, Jie and Coskun, Ayse and Levenhagen, Michael and Rodrigues, Arun},
title = {SST + gem5 = a scalable simulation infrastructure for high performance computing},
year = {2012},
isbn = {9781450315104},
publisher = {ICST (Institute for Computer Sciences, Social-Informatics and Telecommunications Engineering)},
address = {Brussels, BEL},
booktitle = {Proceedings of the 5th International ICST Conference on Simulation Tools and Techniques},
pages = {196–201},
numpages = {6},
location = {Desenzano del Garda, Italy},
series = {SIMUTOOLS '12}
}

@INPROCEEDINGS{hopp,
  author={Li, Haifeng and Liu, Ke and Liang, Ting and Li, Zuojun and Lu, Tianyue and Yuan, Hui and Xia, Yinben and Bao, Yungang and Chen, Mingyu and Shan, Yizhou},
  booktitle={2023 IEEE International Symposium on High-Performance Computer Architecture (HPCA)}, 
  title="{HoPP: Hardware-Software Co-Designed Page Prefetching for Disaggregated Memory}", 
  year={2023},
  volume={},
  number={},
  pages={1168-1181},
  doi={10.1109/HPCA56546.2023.10070986}}

@misc{dassharmax,
      title={An Introduction to the Compute Express Link (CXL) Interconnect}, 
      author={Debendra Das Sharma and Robert Blankenship and Daniel S. Berger},
      year={2024},
      eprint={2306.11227},
      archivePrefix={arXiv},
      primaryClass={cs.AR}
}

@inproceedings{pond,
author = {Li, Huaicheng and Berger, Daniel S. and Hsu, Lisa and Ernst, Daniel and Zardoshti, Pantea and Novakovic, Stanko and Shah, Monish and Rajadnya, Samir and Lee, Scott and Agarwal, Ishwar and Hill, Mark D. and Fontoura, Marcus and Bianchini, Ricardo},
title = {Pond: CXL-Based Memory Pooling Systems for Cloud Platforms},
year = {2023},
isbn = {9781450399166},
publisher = {Association for Computing Machinery},
address = {New York, NY, USA},
url = {https://doi.org/10.1145/3575693.3578835},
doi = {10.1145/3575693.3578835},
booktitle = {Proceedings of the 28th ACM International Conference on Architectural Support for Programming Languages and Operating Systems, Volume 2},
pages = {574–587},
numpages = {14},
location = {Vancouver, BC, Canada},
series = {ASPLOS 2023}
}

@book{og-dismem,
  title={Issues in the implementation of a remote memory paging system},
  author={Felten, Edward W and Zahorjan, John},
  year={1991},
  publisher={University of Washington, Department of Computer Science and Engineering}
}

@INPROCEEDINGS{cxl-main,
  author={Sharma, Debendra Das},
  booktitle={2022 IEEE Symposium on High-Performance Interconnects (HOTI)}, 
  title={Compute Express Link®: An open industry-standard interconnect enabling heterogeneous data-centric computing}, 
  year={2022},
  volume={},
  number={},
  pages={5-12},
  doi={10.1109/HOTI55740.2022.00017}}

@article{mem-disaggregation,
author = {Aguilera, Marcos K. and Amaro, Emmanuel and Amit, Nadav and Hunhoff, Erika and Yelam, Anil and Zellweger, Gerd},
title = {Memory disaggregation: why now and what are the challenges},
year = {2023},
issue_date = {June 2023},
publisher = {Association for Computing Machinery},
address = {New York, NY, USA},
volume = {57},
number = {1},
issn = {0163-5980},
url = {https://doi.org/10.1145/3606557.3606563},
doi = {10.1145/3606557.3606563},
journal = {SIGOPS Oper. Syst. Rev.},
month = {jun},
pages = {38–46},
numpages = {9}
}

@misc{cxl-3.1-spec,
  title="{CXL® 3.1 Specification}",
  url={https://computeexpresslink.org/cxl-specification/"},
}

@misc{cxl-4-spec,
  title="{CXL® 4.0 Specification}",
  url={https://computeexpresslink.org/cxl-specification/"},
}

@techreport{genz,
    author = {Gen-Z Interconnect},
    title = {Gen-z zmmu and memory interleave,},
    institution = {Gen-Z Consortium},
    Month = {July}, 
    year = {2017},
    url = {https://genzconsortium.org/wp-content/uploads/2018/05/Gen-Z-MMUand-Memory-Interleave.pdf}
}

@article{cheri-flexible,
author = {Stark, Samuel W. and Markettos, A. Theodore and Moore, Simon W.},
title = {How Flexible is CXL's Memory Protection? Replacing a sledgehammer with a scalpel},
year = {2023},
issue_date = {May/June 2023},
publisher = {Association for Computing Machinery},
address = {New York, NY, USA},
volume = {21},
number = {3},
issn = {1542-7730},
url = {https://doi.org/10.1145/3606014},
doi = {10.1145/3606014},
journal = {Queue},
month = jul,
pages = {54–64},
numpages = {11}
}

@inproceedings{kvstore,
  title={Tigon: A distributed database for a CXL pod},
  author={Huang, Yibo and Chen, Haowei and Ni, Newton and Chidambaram, Vijay and Tang, Dixin and Witchel, Emmett and Zhu, Zhiting and Jia, Zhipeng},
  booktitle={19th USENIX Symposium on Operating Systems Design and Implementation (OSDI 25), Boston, MA},
  year={2025}
}

@misc{cgroups,
    title = "{cgroup\_namespaces(7) — Linux manual page}",
    institution = "{Linux Kernel}"
}

@INPROCEEDINGS{deact,
  author={Kommareddy, Vamsee Reddy and Hughes, Clayton and Hammond, Simon David and Awad, Amro},
  booktitle={2021 IEEE International Symposium on High-Performance Computer Architecture (HPCA)}, 
  title={DeACT: Architecture-Aware Virtual Memory Support for Fabric Attached Memory Systems}, 
  year={2021},
  volume={},
  number={},
  pages={453-466},
  doi={10.1109/HPCA51647.2021.00046}}

@INPROCEEDINGS{minerva,
  author={Alwadi, Mazen and Wang, Rujia and Mohaisen, David and Hughes, Clayton and Hammond, Simon David and Awad, Amro},
  booktitle={2022 IEEE International Parallel and Distributed Processing Symposium (IPDPS)},
  title={Minerva: Rethinking Secure Architectures for the Era of Fabric-Attached Memory Architectures},
  year={2022},
  volume={},
  number={},
  pages={258-268},
  doi={10.1109/IPDPS53621.2022.00033}}

@book{osbook,
  title={Operating systems: Three easy pieces},
  author={Arpaci-Dusseau, Remzi H and Arpaci-Dusseau, Andrea C},
  year={2018},
  publisher={Arpaci-Dusseau Books, LLC}
}

@INPROCEEDINGS{ayaz-ipdps,
  author={Akram, Ayaz and Giannakou, Anna and Akella, Venkatesh and Lowe-Power, Jason and Peisert, Sean},
  booktitle={2021 IEEE International Parallel and Distributed Processing Symposium (IPDPS)}, 
  title={Performance Analysis of Scientific Computing Workloads on General Purpose TEEs}, 
  year={2021},
  volume={},
  number={},
  pages={1066-1076},
  doi={10.1109/IPDPS49936.2021.00115}}

@inproceedings {legoos,
author = {Yizhou Shan and Yutong Huang and Yilun Chen and Yiying Zhang},
title = {{LegoOS}: A Disseminated, Distributed {OS} for Hardware Resource Disaggregation},
booktitle = {13th USENIX Symposium on Operating Systems Design and Implementation (OSDI 18)},
year = {2018},
isbn = {978-1-939133-08-3},
address = {Carlsbad, CA},
pages = {69--87},
url = {https://www.usenix.org/conference/osdi18/presentation/shan},
publisher = {USENIX Association},
month = oct
}

@inproceedings {elasticlave,
author = {Jason Zhijingcheng Yu and Shweta Shinde and Trevor E. Carlson and Prateek Saxena},
title = {Elasticlave: An Efficient Memory Model for Enclaves},
booktitle = {31st USENIX Security Symposium (USENIX Security 22)},
year = {2022},
isbn = {978-1-939133-31-1},
address = {Boston, MA},
pages = {4111--4128},
url = {https://www.usenix.org/conference/usenixsecurity22/presentation/yu-jason},
publisher = {USENIX Association},
month = aug
}

@inproceedings{cheri-paper,
author = {Woodruff, Jonathan and Watson, Robert N.M. and Chisnall, David and Moore, Simon W. and Anderson, Jonathan and Davis, Brooks and Laurie, Ben and Neumann, Peter G. and Norton, Robert and Roe, Michael},
title = {The CHERI capability model: revisiting RISC in an age of risk},
year = {2014},
isbn = {9781479943944},
publisher = {IEEE Press},
booktitle = {Proceeding of the 41st Annual International Symposium on Computer Architecuture},
pages = {457–468},
numpages = {12},
location = {Minneapolis, Minnesota, USA},
series = {ISCA '14}
}

@INPROCEEDINGS{cheri-sp,
  author={Watson, Robert N.M. and Woodruff, Jonathan and Neumann, Peter G. and Moore, Simon W. and Anderson, Jonathan and Chisnall, David and Dave, Nirav and Davis, Brooks and Gudka, Khilan and Laurie, Ben and Murdoch, Steven J. and Norton, Robert and Roe, Michael and Son, Stacey and Vadera, Munraj},
  booktitle={2015 IEEE Symposium on Security and Privacy}, 
  title={CHERI: A Hybrid Capability-System Architecture for Scalable Software Compartmentalization}, 
  year={2015},
  volume={},
  number={},
  pages={20-37},
  doi={10.1109/SP.2015.9}}

@inproceedings{bordercontrol,
author = {Olson, Lena E. and Power, Jason and Hill, Mark D. and Wood, David A.},
title = {Border control: sandboxing accelerators},
year = {2015},
isbn = {9781450340342},
publisher = {Association for Computing Machinery},
address = {New York, NY, USA},
url = {https://doi.org/10.1145/2830772.2830819},
doi = {10.1145/2830772.2830819},
booktitle = {Proceedings of the 48th International Symposium on Microarchitecture},
pages = {470–481},
numpages = {12},
location = {Waikiki, Hawaii},
series = {MICRO-48}
}

@inproceedings{meltdown,
 author = {Moritz Lipp and Michael Schwarz and Daniel Gruss and Thomas Prescher and Werner Haas and Anders Fogh and Jann Horn and Stefan Mangard and Paul Kocher and Daniel Genkin and Yuval Yarom and Mike Hamburg},
 title = {Meltdown: Reading Kernel Memory from User Space},
 booktitle = {27th {USENIX} Security Symposium ({USENIX} Security 18)},
 year = {2018},
}

@inproceedings{spectre,
 author = {Paul Kocher and Jann Horn and Anders Fogh and and Daniel Genkin and Daniel Gruss and Werner Haas and Mike Hamburg and Moritz Lipp and Stefan Mangard and Thomas Prescher and Michael Schwarz and Yuval Yarom},
 title = {Spectre Attacks: Exploiting Speculative Execution},
 booktitle = {40th IEEE Symposium on Security and Privacy (S\&P'19)},
 year = {2019},
}

@misc{google-zero,
    author="{Google Inc.}",
    title="{Project Zero}",
    url={https://googleprojectzero.blogspot.com/}
}

@misc{gapbs,
      title={The GAP Benchmark Suite}, 
      author={Scott Beamer and Krste Asanović and David Patterson},
      year={2017},
      eprint={1508.03619},
      archivePrefix={arXiv},
      primaryClass={cs.DC},
      url={https://arxiv.org/abs/1508.03619}, 
}

@inproceedings{finemem,
author = {Wang, Xiaoyang and Li, Yongkun and Wu, Kan and Zhu, Wenzhe and Li, Yuqi and Xu, Yinlong},
title = {FineMem: breaking the allocation overhead vs. memory waste dilemma in fine-grained disaggregated memory management},
year = {2025},
isbn = {978-1-939133-47-2},
publisher = {USENIX Association},
address = {USA},
booktitle = {Proceedings of the 19th USENIX Conference on Operating Systems Design and Implementation},
articleno = {4},
numpages = {18},
location = {Boston, MA, USA},
series = {OSDI '25}
}

@article{amdsev,
  title={AMD memory encryption},
  author={Kaplan, David and Powell, Jeremy and Woller, Tom},
  journal={White paper},
  volume={13},
  pages={12},
  year={2016}
}

@techreport{famfs,
    title="{Famfs Shared Memory Filesystem Framework - User Space Repo}",
    author="{MICRON}",
    year={2024},
    url="{https://github.com/cxl-micron-reskit/famfs}"
}

@inproceedings{mondrix,
author = {Witchel, Emmett and Rhee, Junghwan and Asanovi\'{c}, Krste},
title = {Mondrix: memory isolation for linux using mondriaan memory protection},
year = {2005},
isbn = {1595930795},
publisher = {Association for Computing Machinery},
address = {New York, NY, USA},
url = {https://doi.org/10.1145/1095810.1095814},
doi = {10.1145/1095810.1095814},
booktitle = {Proceedings of the Twentieth ACM Symposium on Operating Systems Principles},
pages = {31-44},
numpages = {14},
location = {Brighton, United Kingdom},
series = {SOSP '05}
}

@ARTICLE{intel-mpk,
  author={Park, Soyeon and Lee, Sangho and Kim, Taesoo},
  journal={IEEE Security \& Privacy}, 
  title={Memory Protection Keys: Facts, Key Extension Perspectives, and Discussions}, 
  year={2023},
  volume={21},
  number={3},
  pages={8-15},
  doi={10.1109/MSEC.2023.3250601}}

@techreport{arm-mte,
  author="{ARM}",
  title="{Introduction to the MEmory Tagging Extension}",
  url={https://developer.arm.com/documentation/108035/0100/Introduction-to-the-Memory-Tagging-Extension},
  }

@misc{hdcs-paper,
    title={Memory Sharing with CXL: Hardware and Software Design Approaches},
    author={Sunita Jain and Nagaradhesh Yeleswarapu and Hasan Al Maruf and Rita Gupta},
    year={2024},
    eprint={2404.03245},
    archivePrefix={arXiv},
    primaryClass={cs.ET}
}

@misc{shmem-linux,
  title="{shm overview(7) - Linux manual page}",
  institution = "{Linux Kernel}",
  year={},
  url={https://man7.org/linux/man-pages/man7/shm_overview.7.html}
}

@misc{linux-pasid,
  title="{Shared Virtual Addressing (SVA) with ENQCMD}",
  institution = "{Linux Kernel}",
  year={},
  url={https://www.kernel.org/doc/html/next/x86/sva.html}
}

@misc{intel-pasid,
  title="{Process Address Space ID (PASID)}",
  author="{Intel}",
  url={https://www.intel.com/content/www/us/en/docs/programmable/683059/22-4-9-0-0/process-address-space-id-pasid.html}
}

@article{lampson,
author = {Lampson, Butler W.},
title = {Protection},
year = {1974},
issue_date = {January 1974},
publisher = {Association for Computing Machinery},
address = {New York, NY, USA},
volume = {8},
number = {1},
issn = {0163-5980},
url = {https://doi.org/10.1145/775265.775268},
doi = {10.1145/775265.775268},
journal = {SIGOPS Oper. Syst. Rev.},
month = jan,
pages = {18–24},
numpages = {7}
}

@ARTICLE{polp,
  author={Saltzer, J.H. and Schroeder, M.D.},
  journal={Proceedings of the IEEE}, 
  title={The protection of information in computer systems}, 
  year={1975},
  volume={63},
  number={9},
  pages={1278-1308},
  doi={10.1109/PROC.1975.9939}}

@article{rowhammer,
author = {Kim, Yoongu and Daly, Ross and Kim, Jeremie and Fallin, Chris and Lee, Ji Hye and Lee, Donghyuk and Wilkerson, Chris and Lai, Konrad and Mutlu, Onur},
title = {Flipping Bits in Memory without Accessing Them: An Experimental Study of DRAM Disturbance Errors},
year = {2014},
issue_date = {June 2014},
publisher = {Association for Computing Machinery},
address = {New York, NY, USA},
volume = {42},
number = {3},
issn = {0163-5964},
url = {https://doi.org/10.1145/2678373.2665726},
doi = {10.1145/2678373.2665726},
journal = {SIGARCH Comput. Archit. News},
month = {jun},
pages = {361–372},
numpages = {12}
}

@misc{intel-patent,
  title="{Distributed row hammer tracking}",
  author={Bains, Kuljit S. and Halbert, John B.},
  year={2012},
  publisher={Intel Corporation},
  note="{US Patent US20140095780A1}"
}

@techreport{intel-sgx,
  title="{Intel Software Guard Extensions (Intel SGX)}",
  institution="{Intel Corporation}",
  year={2015},
  url={https://www.intel.com/content/dam/develop/external/us/en/documents/332680-001-720907.pdf},
  }

@techreport{intel-tdx,
  title="{Intel\textregistered\ Trust Domain Extensions (TDX) White Paper}",
  institution="{Intel Corporation}",
  type         = {White Paper},
  year={2022},
  url={https://cdrdv2-public.intel.com/690419/TDX-Whitepaper-February2022.pdf},
  }

@techreport{intel-5-level,
  title="{5-Level Paging and 5-Level EPT}",
  institution="{Intel Corporation}",
  type         = {White Paper},
  year={2017},
  url={https://cdrdv2-public.intel.com/671442/5-level-paging-white-paper.pdf},
  }

@misc{riscv-sv57,
  title="{Virtual Memory Layout on RISC-V Linux}",
  author={Alexandre Ghiti},
  institution = "{Linux Kernel}",
  year={2021},
  url={https://www.kernel.org/doc/html/v6.6/riscv/vm-layout.html},
  }

@misc{uio-kernel,
  title="{The Userspace I/O HOWTO}",
  author={Hans-J\"urgen Koch},
  institution = "{Linux Kernel}",
  year={2006},
  url={https://www.kernel.org/doc/html/v4.14/driver-api/uio-howto.html},
  }

@misc{lam-kernel,
  title="{Support for Intel's Linear Address Masking}",
  author="{Jonathan Corbet}",
  year={2022},
  url={https://lwn.net/Articles/902094/}  
}

@article{hwimpl,
  title={FPGA implementation of an HMAC processor based on the SHA-2 family of hash functions},
  author={Juliato, Marcio and Gebotys, Catherine},
  journal={University of Waterloo, Tech. Rep},
  year={2011}
}

@misc{hmac,
  author       = {Hugo Krawczyk and Mihir Bellare and Ran Canetti},
  title        = {{HMAC}: Keyed-Hashing for Message Authentication},
  howpublished = {RFC 2104, Informational},
  month        = feb,
  year         = {1997},
  url          = {https://datatracker.ietf.org/doc/rfc2104/}
}

@techreport{aes,
  author       = {Joan Daemen and Vincent Rijmen},
  title        = {The Rijndael Block Cipher},
  institution  = {NIST Computer Security Resource Center},
  type         = {AES Submission Document},
  month        = jun,
  year         = {1998},
  url          = {https://csrc.nist.gov/CSRC/media/Projects/Cryptographic-Standards-and-Guidelines/documents/aes-development/rijndael-ammended.pdf}
}

@misc{specific,
  author       = "{Nyström}",
  title        = "{Identifiers and Test Vectors for HMAC‑SHA‑224, HMAC‑SHA‑256, HMAC‑SHA‑384, and HMAC‑SHA‑512}",
  howpublished = {RFC 4231},
  year         = {2005},
  url          = {https://datatracker.ietf.org/doc/rfc4231/}
}

@inproceedings{tocttou,
  author    = {Jinpeng Wei and Calton Pu},
  title     = {TOCTTOU Vulnerabilities in UNIX-Style File Systems: An Anatomical Study},
  booktitle = {Proceedings of the 14th USENIX Security Symposium},
  year      = {2005},
  pages     = {225--239}
}

@INPROCEEDINGS{data-enclaves,
  author={Xu, Yuanchao and Pangia, James and Ye, Chencheng and Solihin, Yan and Shen, Xipeng},
  booktitle={2024 IEEE International Symposium on High-Performance Computer Architecture (HPCA)}, 
  title={Data Enclave: A Data-Centric Trusted Execution Environment}, 
  year={2024},
  volume={},
  number={},
  pages={218-232},
  doi={10.1109/HPCA57654.2024.00026}}

@inproceedings{micro-cxl,
author = {Li, Chuanhan and Zhao, Jishen and Xu, Yuanchao},
title = {Efficient Security Support for CXL Memory through Adaptive Incremental Offloaded (Re-)Encryption},
year = {2025},
isbn = {9798400715730},
publisher = {Association for Computing Machinery},
address = {New York, NY, USA},
url = {https://doi.org/10.1145/3725843.3756119},
doi = {10.1145/3725843.3756119},
booktitle = {Proceedings of the 58th IEEE/ACM International Symposium on Microarchitecture},
pages = {1102–1116},
numpages = {15},
location = {
},
series = {MICRO '25}
}

@article{taco-oram-cxl,
author = {Choi, Kwanghoon and Kim, Igjae and Lee, Sunho and Huh, Jaehyuk},
title = {ShieldCXL: A Practical Obliviousness Support with Sealed CXL Memory},
year = {2025},
issue_date = {March 2025},
publisher = {Association for Computing Machinery},
address = {New York, NY, USA},
volume = {22},
number = {1},
issn = {1544-3566},
url = {https://doi.org/10.1145/3703354},
doi = {10.1145/3703354},
journal = {ACM Trans. Archit. Code Optim.},
month = mar,
articleno = {13},
numpages = {25}
}

@ARTICLE{cheri-rackscale,
  author={Bresniker, Kirk M. and Faraboschi, Paolo and Mendelson, Avi and Milojicic, Dejan and Roscoe, Timothy and Watson, Robert N.M.},
  journal={Computer}, 
  title={Rack-Scale Capabilities: Fine-Grained Protection for Large-Scale Memories}, 
  year={2019},
  volume={52},
  number={2},
  pages={52-62},
  doi={10.1109/MC.2018.2888769}}

@ARTICLE{cosm,
  author={Sundaravarathan, Vignesh and Reisslein, Martin and Thyagaturu, Akhilesh S. and Ross, Nick and Singh Kalsi, Gurpreet and Howard, Jason and Kaisrlik, Jan and Matwiejczyk, Bartosz and Landowski, Marek M. and Dorozynski, Piotr and Vrsalovic, Harvey and Tayal, Sanjaya},
  journal={IEEE Access}, 
  title={Controlled Shared Memory (COSM) Isolation: Design and Testbed Evaluation}, 
  year={2025},
  volume={13},
  number={},
  pages={77893-77917},
  doi={10.1109/ACCESS.2025.3564391}}

@misc{cxl-graphs,
author={CXL},
title="{Fueling AI and HPC Workloads with CXL at SC25}",
year={2025},
url={https://computeexpresslink.org/blog/fueling-ai-and-hpc-workloads-with-cxl-at-sc25-4233/}
}

@phdthesis{ayaz-thesis,
  title={Hardware/Software Co-Design for Secure High Performance Computing Systems},
  author={Akram, Ayaz},
  year={2023},
  school={University of California, Davis}
}

@misc{cxl-clustersim,
      title="{CXL-ClusterSim: Modeling CXL-based Disaggregated Memory Cluster for Pooling and Sharing using gem5 and SST}", 
      author={Kaustav Goswami and Maryam Babaie and Hoa Nguyen and Venkatesh Akella and Jason Lowe-Power},
      year={2026},
      eprint={2605.27745},
      archivePrefix={arXiv},
      primaryClass={cs.AR},
      url={https://arxiv.org/abs/2605.27745}, 
}

\section{Address Translation and Access Control in CXL}
    \label{app:cxl-translation}
    
        CXL defines \textit{hosts} and \textit{devices}.
        Hosts are compute nodes with or without local memory.
        The devices on the other hand, are memory devices.
        There is a global address space, where devices are mapped into.
        Hosts may choose to map a part of their own address space into the global address space.
        The global address space is one contiguous address space.
        Until CXL 2.0, there was no notion of sharing memory ranges among hosts.
        To facilitate memory sharing, additional hardware is added.
        The fabric manager (FM) is responsible for binding hosts and devices to the global address space.
        A Global Fabric Attached Memory (G-FAM) is a collection of hosts and devices in one single contiguous global physical address space.
        CXL switches are responsible for routing a given request to its corresponding device.
    
        Hosts can have both local and remote memory ranges.
        Local ranges are not exposed to the G-FAM, therefore, not appearing in the global address space.
        The remote memory addresses reside outside the host.
        Note that both of these addresses are physical addresses.
        To distinguish between host's physical addresses and a device's physical address, the terminology defines host physical address (HPA) and device physical address (DPA).
        A host is assigned a source port ID (SPID) and a device is assigned a device port ID (DPID).
    
        New tables are added to the CXL switch and the G-FAM device to enable sharing.
        Fabric address segment table (FAST) and the Interleave DPID table (IDT) are added to a CXL switch.
        On the other hand, G-FAM devices (GFD) decoder table and the SPID Access
        Table (SAT) are added to the G-FAM device.
        In CXL terminology, devices are referred to as dynamic capacity devices (DCD).
        A DCD is divided into multiple regions called DCD region of size 256 MiB~\cite{cxl-3.1-spec}.
        Access permissions for a host is determined at the granularity of a DCD region.
    
        \begin{figure}[t]
            \centering
            \includegraphics[width=1\columnwidth]{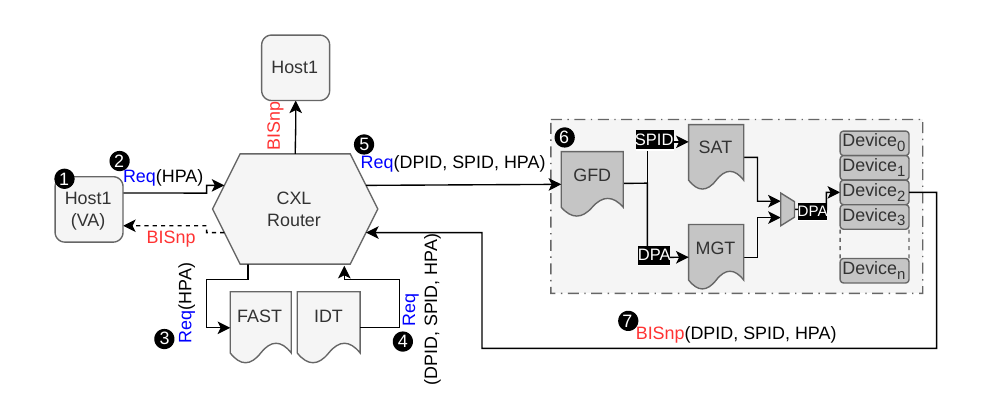}
            \caption{Address translation and access control in CXL.
                    The figure defines a G-FAM system with \textit{two} hosts and \textit{n} devices.}
            \label{fig:addr-translation}
        \end{figure}

        Figure~\ref{fig:addr-translation} shows the journey of a virtual address generated at a host to its corresponding device.
        Action~\encircle{1} in the figure shows the generation of virtual address (VA) at the host.
        The host translates this VA to a host physical address (HPA) using the host's MMU/TLB and send the HPA as a request address on a CXL flit to the outgoing CXL downstream port.
        
        The first component a CXL flit encounters is a CXL router.
        The router is responsible for appending the SPID and the DPID of the address into the request flit.
        Action~\encircle{3} shows that forwarding of the request flit to the FAST table to determine the DPID.
        The request is further sent to the IDT to determine if the address is interleaved into multiple GFDs.
        If a valid destination does not exist for the DPID, the host is notified and the flit is dropped.
        The appended request can be seen in Action~\encircle{4}.
        The CXL router then forwards this request to the GFD (Action~\encircle{5}).
    
        The tables for one set of G-FAM is responsible for checking access permission for a given address on a flit for the corresponding DPID and the SPID.
        The incoming request flit address is looked up in the GFD table (Action~\encircle{6}).
        Once the correct GFD is decoded, the request is then forwarded to the SAT.
        The GFD is also responsible for translating the HPA to its corresponding device physical address (DPA).
        The SAT table hosts access permission for a memory group.
        The table is querried with the SPID to determine whether the given host has permissions for a given memory group.

        In case the flit is a write on a HPA, a Back-Invalidate Snoop (BISnp) is sent back to all the hosts sharing that HPA (Action~\encircle{7}).

\end{document}